\documentclass[twocolumn,letterpaper,aps,prc,superscriptaddress,nofootinbib,showpacs,floatfix]{revtex4-1}

\usepackage{graphicx}	% Include figure files
\usepackage{xspace}     % Include xspace
\usepackage{multirow} 

\newcommand{\pt}{\mbox{$p_T$}\xspace}

\newcommand{\raa}{\mbox{$R_{AA}$}\xspace}

\newcommand{\Npart}{\mbox{$N_{\rm part}$}\xspace}
\newcommand{\Ncoll}{\mbox{$N_{\rm coll}$}\xspace}

\newcommand{\sqsn}{\mbox{$\sqrt{s_{_{NN}}}$}\xspace}
\newcommand{\jpsi}{\mbox{$J/\psi$}\xspace}
\newcommand{\psip}{\mbox{$\psi^{\prime}$}\xspace}
\newcommand{\mumu}{\mbox{$\mu^{+}\mu^{-}$}\xspace}
\newcommand{\pp}{\mbox{$p$$+$$p$}\xspace}
\newcommand{\auau}{\mbox{Au$+$Au}\xspace}
\newcommand{\cucu}{\mbox{Cu+Cu}\xspace}
\newcommand{\cuau}{\mbox{Cu+Au}\xspace}
\newcommand{\dau}{\mbox{$d$$+$Au}\xspace}
\newcommand{\ppb}{\mbox{$p$$+$Pb}\xspace}
\newcommand{\pbpb}{\mbox{Pb+Pb}\xspace}
\newcommand{\inin}{\mbox{In+In}\xspace}
\newcommand{\su}{\mbox{S+U}\xspace}
\newcommand{\pda}{\mbox{$p(d)$+$A$}\xspace}
\newcommand{\yforward}{\mbox{$1.2$$<$$y$$<$$2.2$}\xspace}
\newcommand{\ybackward}{\mbox{$-2.2$$<$$y$$<$$-1.2$}\xspace}

\begin{document}

%%%%%%%%%%%%%%%%%%%%%%%%%%%%%%%%%%%%%%%%%%%%%%%%% Title of paper

\title{Nuclear matter effects on $J/\psi$ production in
  asymmetric Cu+Au collisions at $\sqrt{s_{_{NN}}}$=200~GeV.}

\newcommand{\abilene}{Abilene Christian University, Abilene, Texas 79699, USA}
\newcommand{\augie}{Department of Physics, Augustana College, Sioux Falls, South Dakota 57197, USA}
\newcommand{\banaras}{Department of Physics, Banaras Hindu University, Varanasi 221005, India}
\newcommand{\barc}{Bhabha Atomic Research Centre, Bombay 400 085, India}
\newcommand{\baruch}{Baruch College, City University of New York, New York, New York, 10010 USA}
\newcommand{\bnlcoll}{Collider-Accelerator Department, Brookhaven National Laboratory, Upton, New York 11973-5000, USA}
\newcommand{\bnlphys}{Physics Department, Brookhaven National Laboratory, Upton, New York 11973-5000, USA}
\newcommand{\caucr}{University of California - Riverside, Riverside, California 92521, USA}
\newcommand{\charlesczech}{Charles University, Ovocn\'{y} trh 5, Praha 1, 116 36, Prague, Czech Republic}
\newcommand{\chonbuk}{Chonbuk National University, Jeonju, 561-756, Korea}
\newcommand{\ciae}{Science and Technology on Nuclear Data Laboratory, China Institute of Atomic Energy, Beijing 102413, P.~R.~China}
\newcommand{\cns}{Center for Nuclear Study, Graduate School of Science, University of Tokyo, 7-3-1 Hongo, Bunkyo, Tokyo 113-0033, Japan}
\newcommand{\colorado}{University of Colorado, Boulder, Colorado 80309, USA}
\newcommand{\columbia}{Columbia University, New York, New York 10027 and Nevis Laboratories, Irvington, New York 10533, USA}
\newcommand{\czechtech}{Czech Technical University, Zikova 4, 166 36 Prague 6, Czech Republic}
\newcommand{\elte}{ELTE, E{\"o}tv{\"o}s Lor{\'a}nd University, H - 1117 Budapest, P{\'a}zm{\'a}ny P. s. 1/A, Hungary}
\newcommand{\ewha}{Ewha Womans University, Seoul 120-750, Korea}
\newcommand{\fsu}{Florida State University, Tallahassee, Florida 32306, USA}
\newcommand{\gsu}{Georgia State University, Atlanta, Georgia 30303, USA}
\newcommand{\hanyang}{Hanyang University, Seoul 133-792, Korea}
\newcommand{\hiroshima}{Hiroshima University, Kagamiyama, Higashi-Hiroshima 739-8526, Japan}
\newcommand{\ihepprot}{IHEP Protvino, State Research Center of Russian Federation, Institute for High Energy Physics, Protvino, 142281, Russia}
\newcommand{\illuiuc}{University of Illinois at Urbana-Champaign, Urbana, Illinois 61801, USA}
\newcommand{\inrras}{Institute for Nuclear Research of the Russian Academy of Sciences, prospekt 60-letiya Oktyabrya 7a, Moscow 117312, Russia}
\newcommand{\instpasczech}{Institute of Physics, Academy of Sciences of the Czech Republic, Na Slovance 2, 182 21 Prague 8, Czech Republic}
\newcommand{\isu}{Iowa State University, Ames, Iowa 50011, USA}
\newcommand{\jaea}{Advanced Science Research Center, Japan Atomic Energy Agency, 2-4 Shirakata Shirane, Tokai-mura, Naka-gun, Ibaraki-ken 319-1195, Japan}
\newcommand{\jyvaskyla}{Helsinki Institute of Physics and University of Jyv{\"a}skyl{\"a}, P.O.Box 35, FI-40014 Jyv{\"a}skyl{\"a}, Finland}
\newcommand{\kek}{KEK, High Energy Accelerator Research Organization, Tsukuba, Ibaraki 305-0801, Japan}
\newcommand{\korea}{Korea University, Seoul, 136-701, Korea}
\newcommand{\kurchatov}{Russian Research Center ``Kurchatov Institute", Moscow, 123098 Russia}
\newcommand{\kyoto}{Kyoto University, Kyoto 606-8502, Japan}
\newcommand{\labllr}{Laboratoire Leprince-Ringuet, Ecole Polytechnique, CNRS-IN2P3, Route de Saclay, F-91128, Palaiseau, France}
\newcommand{\lahorelums}{Physics Department, Lahore University of Management Sciences, Lahore, Pakistan}
\newcommand{\lawllnl}{Lawrence Livermore National Laboratory, Livermore, California 94550, USA}
\newcommand{\losalamos}{Los Alamos National Laboratory, Los Alamos, New Mexico 87545, USA}
\newcommand{\lund}{Department of Physics, Lund University, Box 118, SE-221 00 Lund, Sweden}
\newcommand{\maryland}{University of Maryland, College Park, Maryland 20742, USA}
\newcommand{\mass}{Department of Physics, University of Massachusetts, Amherst, Massachusetts 01003-9337, USA }
\newcommand{\michigan}{Department of Physics, University of Michigan, Ann Arbor, Michigan 48109-1040, USA}
\newcommand{\muhlenberg}{Muhlenberg College, Allentown, Pennsylvania 18104-5586, USA}
\newcommand{\myongji}{Myongji University, Yongin, Kyonggido 449-728, Korea}
\newcommand{\newmex}{University of New Mexico, Albuquerque, New Mexico 87131, USA }
\newcommand{\nmsu}{New Mexico State University, Las Cruces, New Mexico 88003, USA}
\newcommand{\ohio}{Department of Physics and Astronomy, Ohio University, Athens, Ohio 45701, USA}
\newcommand{\ornl}{Oak Ridge National Laboratory, Oak Ridge, Tennessee 37831, USA}
\newcommand{\orsay}{IPN-Orsay, Universite Paris Sud, CNRS-IN2P3, BP1, F-91406, Orsay, France}
\newcommand{\pnpi}{PNPI, Petersburg Nuclear Physics Institute, Gatchina, Leningrad region, 188300, Russia}
\newcommand{\riken}{RIKEN Nishina Center for Accelerator-Based Science, Wako, Saitama 351-0198, Japan}
\newcommand{\rikjrbrc}{RIKEN BNL Research Center, Brookhaven National Laboratory, Upton, New York 11973-5000, USA}
\newcommand{\rikkyo}{Physics Department, Rikkyo University, 3-34-1 Nishi-Ikebukuro, Toshima, Tokyo 171-8501, Japan}
\newcommand{\saispbstu}{Saint Petersburg State Polytechnic University, St. Petersburg, 195251 Russia}
\newcommand{\seoulnat}{Department of Physics and Astronomy, Seoul National University, Seoul, Korea}
\newcommand{\stonybrkc}{Chemistry Department, Stony Brook University, SUNY, Stony Brook, New York 11794-3400, USA}
\newcommand{\stonycrkp}{Department of Physics and Astronomy, Stony Brook University, SUNY, Stony Brook, New York 11794-3800,, USA}
\newcommand{\tenn}{University of Tennessee, Knoxville, Tennessee 37996, USA}
\newcommand{\titech}{Department of Physics, Tokyo Institute of Technology, Oh-okayama, Meguro, Tokyo 152-8551, Japan}
\newcommand{\tsukuba}{Institute of Physics, University of Tsukuba, Tsukuba, Ibaraki 305, Japan}
\newcommand{\vandy}{Vanderbilt University, Nashville, Tennessee 37235, USA}
\newcommand{\weizmann}{Weizmann Institute, Rehovot 76100, Israel}
\newcommand{\wigner}{Institute for Particle and Nuclear Physics, Wigner Research Centre for Physics, Hungarian Academy of Sciences (Wigner RCP, RMKI) H-1525 Budapest 114, POBox 49, Budapest, Hungary}
\newcommand{\yonsei}{Yonsei University, IPAP, Seoul 120-749, Korea}
\newcommand{\zagreb}{University of Zagreb, Faculty of Science, Department of Physics, Bijeni\v{c}ka 32, HR-10002 Zagreb, Croatia}
\affiliation{\abilene}
\affiliation{\augie}
\affiliation{\banaras}
\affiliation{\barc}
\affiliation{\baruch}
\affiliation{\bnlcoll}
\affiliation{\bnlphys}
\affiliation{\caucr}
\affiliation{\charlesczech}
\affiliation{\chonbuk}
\affiliation{\ciae}
\affiliation{\cns}
\affiliation{\colorado}
\affiliation{\columbia}
\affiliation{\czechtech}
\affiliation{\elte}
\affiliation{\ewha}
\affiliation{\fsu}
\affiliation{\gsu}
\affiliation{\hanyang}
\affiliation{\hiroshima}
\affiliation{\ihepprot}
\affiliation{\illuiuc}
\affiliation{\inrras}
\affiliation{\instpasczech}
\affiliation{\isu}
\affiliation{\jaea}
\affiliation{\jyvaskyla}
\affiliation{\kek}
\affiliation{\korea}
\affiliation{\kurchatov}
\affiliation{\kyoto}
\affiliation{\labllr}
\affiliation{\lahorelums}
\affiliation{\lawllnl}
\affiliation{\losalamos}
\affiliation{\lund}
\affiliation{\maryland}
\affiliation{\mass}
\affiliation{\michigan}
\affiliation{\muhlenberg}
\affiliation{\myongji}
\affiliation{\newmex}
\affiliation{\nmsu}
\affiliation{\ohio}
\affiliation{\ornl}
\affiliation{\orsay}
\affiliation{\pnpi}
\affiliation{\riken}
\affiliation{\rikjrbrc}
\affiliation{\rikkyo}
\affiliation{\saispbstu}
\affiliation{\seoulnat}
\affiliation{\stonybrkc}
\affiliation{\stonycrkp}
\affiliation{\tenn}
\affiliation{\titech}
\affiliation{\tsukuba}
\affiliation{\vandy}
\affiliation{\weizmann}
\affiliation{\wigner}
\affiliation{\yonsei}
\affiliation{\zagreb}
\author{C.~Aidala} \affiliation{\losalamos} \affiliation{\michigan}
\author{N.N.~Ajitanand} \affiliation{\stonybrkc}
\author{Y.~Akiba} \affiliation{\riken} \affiliation{\rikjrbrc}
\author{R.~Akimoto} \affiliation{\cns}
\author{J.~Alexander} \affiliation{\stonybrkc}
\author{K.~Aoki} \affiliation{\riken}
\author{N.~Apadula} \affiliation{\stonycrkp}
\author{H.~Asano} \affiliation{\kyoto} \affiliation{\riken}
\author{E.T.~Atomssa} \affiliation{\stonycrkp}
\author{T.C.~Awes} \affiliation{\ornl}
% ...
\collaboration{PHENIX Collaboration} \noaffiliation

%------------------------------------------------------------------------------|

\date{\today}

\begin{abstract}

%\linenumbers

We report on $J/\psi$ production from asymmetric Cu+Au heavy-ion collisions 
at $\sqrt{s_{_{NN}}}$=200~GeV at the Relativistic Heavy Ion Collider at both 
forward (Cu-going direction) and backward (Au-going direction) rapidities.  
The nuclear modification of $J/\psi$ yields in Cu$+$Au collisions in the 
Au-going direction is found to be comparable to that in Au$+$Au collisions 
when plotted as a function of the number of participating nucleons. In the 
Cu-going direction, $J/\psi$ production shows a stronger suppression.  This 
difference is comparable in magnitude and has the same sign as the 
difference expected from shadowing effects due to stronger low-$x$ gluon 
suppression in the larger Au nucleus. The relative suppression is opposite 
to that expected from hot nuclear matter dissociation, since a higher 
energy density is expected in the Au-going direction.

\end{abstract}


\maketitle

\section{Introduction}

The long-standing goal of studying the production in high energy heavy ion 
collisions of $c\bar{c}$ bound states, known collectively as charmonium, 
has been to use the modification of their yield as a direct signal of 
deconfinement in the quark gluon plasma 
(QGP)~\cite{Matsui:1986dk,Frawley:2008kk, Brambilla:2010cs}. Practically, 
the study of charmonium has been confined to the two lowest mass vector 
meson states, the strongly bound \jpsi and the much more weakly bound 
\psip. In pursuit of this goal, the production of \jpsi has been studied 
at center of mass energies of \sqsn = 17.3\,GeV in 
\pbpb~\cite{Alessandro:2004ap}, \inin~\cite{Arnaldi:2007zz}, and 
\ppb~\cite{Arnaldi:2010ky,Arnaldi:2009ph} collisions; at 
\sqsn\,=\,19.4\,GeV in \su collisions~\cite{Abreu:1999nn}; at 
\sqsn\,=\,200\,GeV in \pp~\cite{Adare:2011vq}, 
\dau~\cite{Adare:2010fn,Adare:2012qf}, \cucu~\cite{Adare:2008sh} and 
\auau~\cite{Adare:2011yf,Adare:2006ns} collisions; and at 
\sqsn\,=\,2.76--7\,TeV in \pp~\cite{Abelev:2012kr,Aamodt:2011gj}, 
\ppb~\cite{Abelev:2013yxa} and \pbpb~\cite{Abelev:2013ila} collisions. 
Only one heavy ion on heavy ion collision system has asymmetric masses, 
\su at 19.4\,GeV, and that measurement was made at only one rapidity 
(0$<$$y$$<$1).

The studies of \pda collisions at these and other energies were motivated 
by the need to understand cold nuclear matter (CNM) 
effects~\cite{Frawley:2008kk,Brambilla:2010cs}. These are effects that 
modify \jpsi production in a nuclear target in the absence of a QGP, and 
they are found to be very significant at all of these 
energies~\cite{Arnaldi:2010ky,Adare:2010fn,Abelev:2013yxa,Leitch:1999ea,Abt:2008ya,Badier:1983dg,Alessandro:2003pc,Alessandro:2006jt}. 
CNM effects often considered include nuclear modification of the parton 
distributions in nuclei (nPDFs), break up of the \jpsi precursor 
$c\bar{c}$ state in the cold nucleus, nuclear transverse momentum 
broadening in traversing the cold nucleus, and initial state parton energy 
loss~\cite{Frawley:2008kk,Brambilla:2010cs}. It has been hoped that CNM 
effects and hot matter effects can be factorized, so that CNM effects can 
be measured in \pda collisions and accounted for when analyzing heavy ion 
collision data to extract hot matter effects. This has not yet been 
clearly established.

The recent observation of what appears to be collective flow in \ppb 
\cite{Abelev:2012ola,Aad:2013fja,Chatrchyan:2013nka} and \dau 
\cite{Adare:2013piz} collisions has called into question whether CNM 
effects are really isolated from hot matter effects in \pda collisions. 
Evidence that \jpsi production is not modified by hot matter effects in 
\pda collisions comes from the observation~\cite{McGlinchey:2012bp} that 
break up cross sections fitted to shadowing corrected \jpsi data from \pda 
collisions at mid and backward rapidity scale with time spent in the 
nucleus across a broad range of collision energies. This observed scaling 
would presumably be broken if \jpsi production was modified by different 
hot matter effects at different collision energies. However unexpectedly 
strong suppression of the \psip has been observed in both 
\dau~\cite{Adare:2013ezl} and $p$$+$Pb~\cite{Winn:HP2013} collisions, and 
so far this is unexplained. Since feed down from \psip decays contributes 
only 10\% to the \jpsi yield, it is possible that the weakly bound \psip 
is sensitive to hot matter effects in \pda collisions while the inclusive 
\jpsi yield is not.

There are additional data from \pda collisions at lower collision energies 
\cite{Leitch:1999ea,Abt:2008ya,Badier:1983dg,Alessandro:2003pc,Alessandro:2006jt}. 
Taken together with the \pda data sets mentioned above, they cover a broad 
range of rapidities and \sqsn values. To try to shed some light on the 
nature of CNM effects on \jpsi production, these data have been described 
using models containing gluon shadowing/antishadowing plus break up of the 
charmonia precursor state by collisions with 
nucleons~\cite{Lourenco:2008sk,Arnaldi:2009ph,McGlinchey:2012bp} 
and/or models of energy loss in cold nuclear 
matter~\cite{Arleo:2012rs,Arleo:2013zua} or gluon saturation 
models~\cite{Kharzeev:2005}. A broad picture now seems to have emerged. 
The precursor to the fully formed charmonium is a $c\bar{c}$ state, formed 
primarily by gluon fusion, that becomes color neutral and expands to the 
final size of the meson on a time scale of a few tenths of a fm/$c$. When 
the proper time (in the $c\bar{c}$ frame) spent in the target nucleus is 
comparable with the charmonium formation time (which occurs at lower 
energies and at midrapidity, and at higher energies only at backward 
rapidities), the modification is well described by shadowing plus break up 
by nucleons~ \cite{McGlinchey:2012bp}.  When the time spent in the target 
nucleus is shorter than this (which occurs at higher energies, and at 
lower energies only at forward rapidity), the data are well described by 
models of shadowing plus energy loss or gluon 
saturation~\cite{Arleo:2012rs,Arleo:2013zua}. Thus at RHIC energy 
(\sqsn\,=\,200\,GeV) cold nuclear matter effects are believed to result 
from a variety of different mechanisms, and the mixture depends very 
strongly on rapidity.

Hot matter effects and CNM effects are present together in heavy ion 
collisions, and both are important. In \auau collisions at RHIC, for 
example, the addition of hot matter effects increases the suppression of 
the \jpsi by a factor of roughly two over what would be expected if only 
CNM effects were present~\cite{Brambilla:2010cs,Adare:2011yf}. Moreover, 
in asymmetric mass collisions such as Cu+Au the distribution of final 
state energy is a function of rapidity \cite{Chen:2005zy}, as reflected in 
the particle production. Thus hot matter effects will likely not be 
symmetric in rapidity. Cold nuclear matter effects will also be asymmetric 
in rapidity. First, the parton distribution functions are more strongly 
modified in the heavier Au nucleus. Forward rapidity (Cu-going) \jpsi 
production probes gluons at low Bjorken-$x$ ({\it i.e.} low momentum 
fraction)  in the Au nucleus, while in Cu the gluons at high Bjorken-$x$ 
are probed. This is reversed for the backward rapidity (Au-going) \jpsi. 
Second, energy loss and breakup effects will be different in nuclei of 
different mass. In the case where the charmonium is emitted at forward 
rapidity it has a large rapidity relative to the Au nucleus, which it 
crosses in a very short proper time. At the same time, the \jpsi rapidity 
relative to the Cu nucleus is much smaller, and the crossing time is much 
larger. Because the different time scales lead to different mechanisms, 
energy loss effects will depend on the interaction between the charmonium 
precursor state and the Au nucleus, while breakup effects will depend on 
the interaction between the precursor and the Cu nucleus. For charmonium 
emitted at backward rapidity, this will be reversed.  Thus the asymmetry 
in mass between Cu and Au will lead to asymmetric energy loss and breakup 
contributions at forward and backward rapidity. Forward versus backward 
rapidity \jpsi production in asymmetric mass collisions will therefore 
contain different contributions from both hot matter effects and CNM 
effects. There are also simple geometric models separating core-corona 
contributions that would be useful to confront with data in central 
Cu+Au~\cite{Digal:2012}. The comparison of \dau, \auau and \cuau \jpsi 
modifications across rapidities may provide key insight on the balance of 
cold and hot nuclear matter effects, and whether they are truly 
factorizable.

A heavy ion collision system with asymmetric masses, \cuau, was studied 
experimentally for the first time at RHIC in the 2012 run. In this paper 
we present nuclear modification data from the PHENIX experiment on \jpsi 
production in \cuau collisions at \sqsn = 200 GeV at two rapidities, 
\ybackward and \yforward.

\section{PHENIX Detector}

%%%%%%%%%%%%%%%%%%%%%%%%%%%%%%%%%%%%%%%%%%%%%%%%%%%%%%%%%% Fig_1
\begin{figure}[!tb]
    \includegraphics[width=1.0\linewidth]{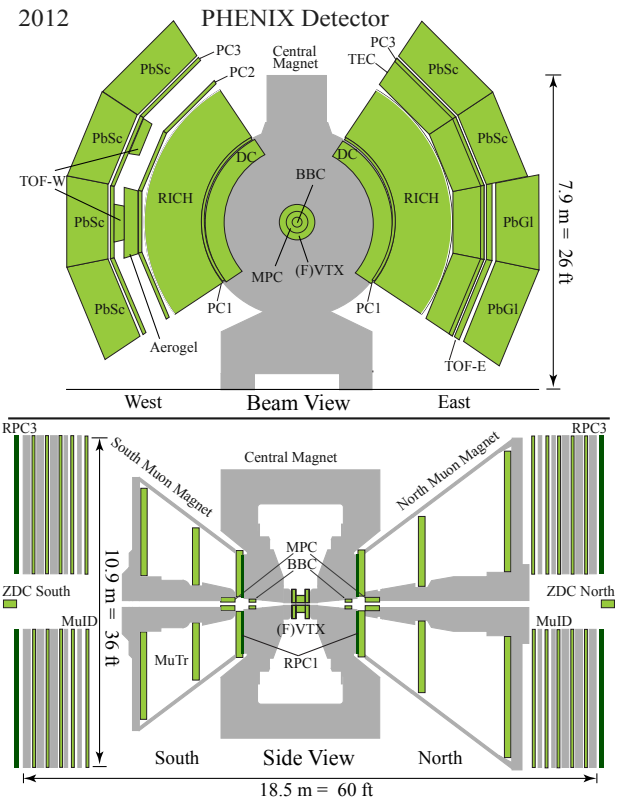}
    \caption{\label{fig:Detector} (Color online)
A schematic side view of the PHENIX detector configuration for
the 2012 run.}
\end{figure}

%%%%%%%%%%%%%%%%%%%%%%%%%%%%%%%%%%%%%%%%%%%%%%%%%%%%%%%%%% Fig_2
\begin{figure*}[!htb]
    \includegraphics[width=0.47\linewidth]{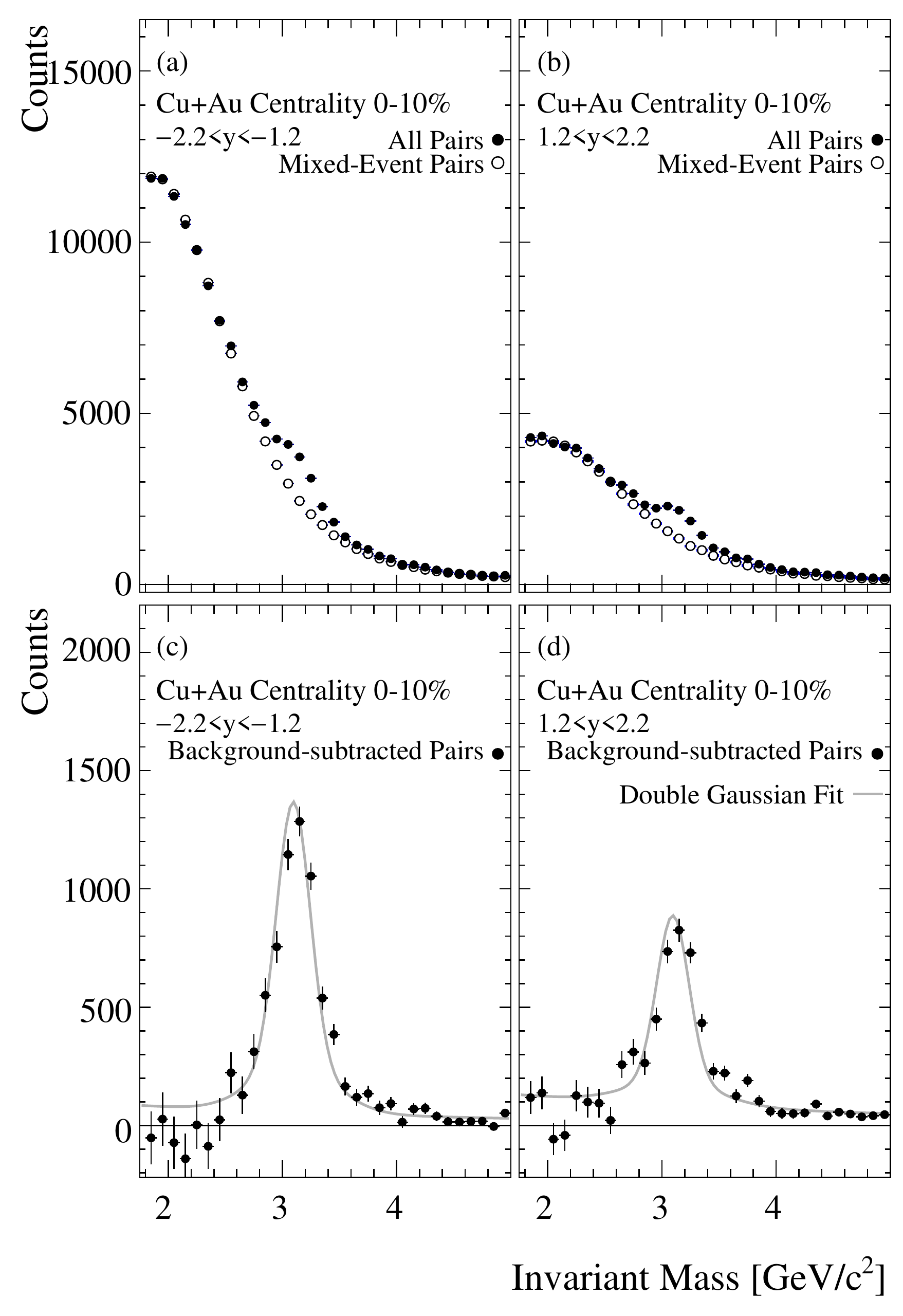}
    \hspace{0.04\linewidth}
    \includegraphics[width=0.47\linewidth]{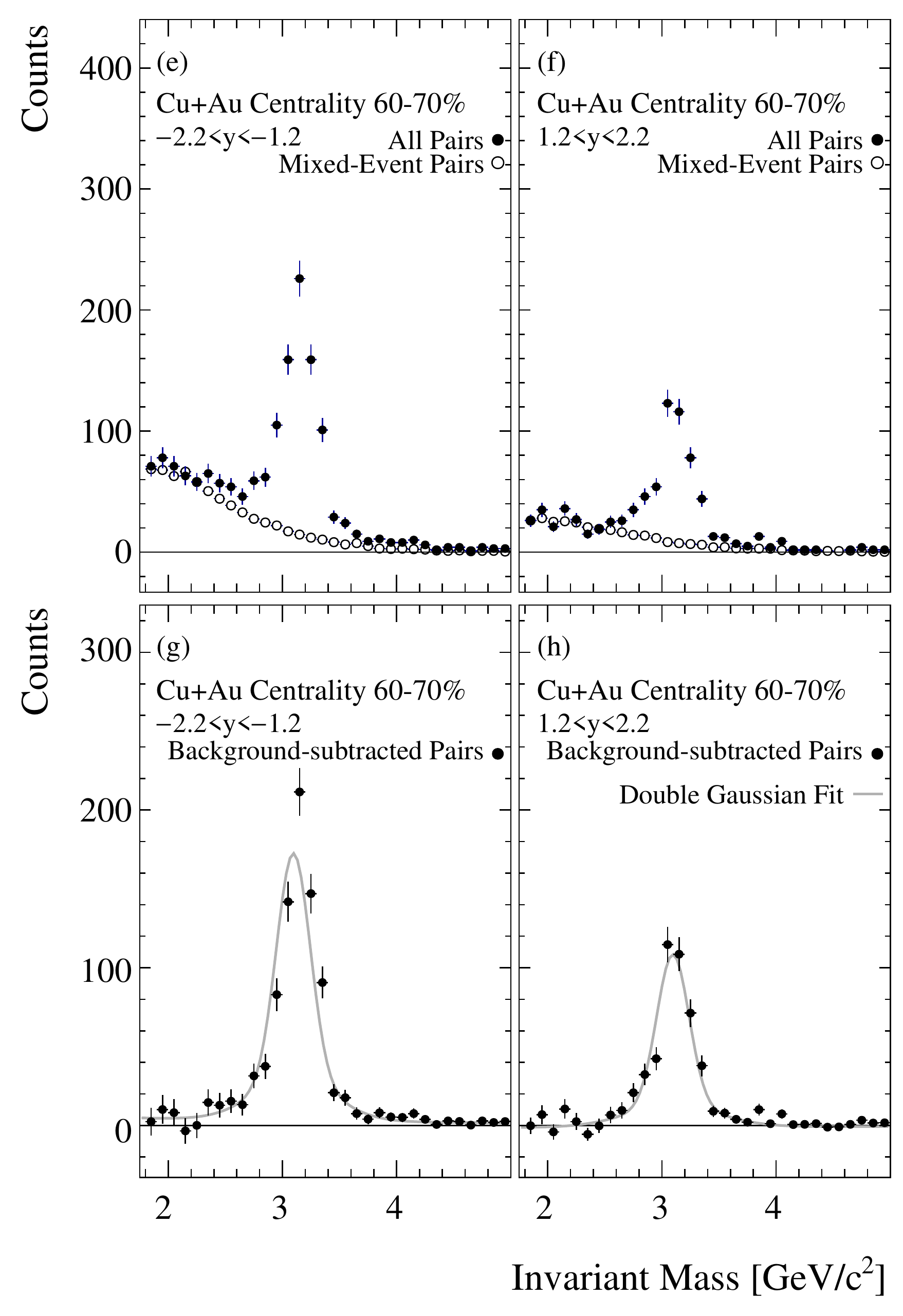}
\caption{\label{fig:RawCounts}
Dimuon invariant mass spectra measured in central 0\%--10\% (left panels 
(a)--(d)) and mid-peripheral (60\%--70\%) (right panels ((e)--(h)) collisions 
integrated over the full \pt range.  In each figure, the top panels 
((a),(b),(e), and (f)) show the distribution of invariant mass, 
reconstructed from all same-event opposite charge-sign pairs (filled 
symbols) and mixed-event pairs (open symbols) in Cu+Au collisions.  The 
lower panels ((c),(d),(g), and (h)) show the combinatorial background 
subtracted pairs from the upper panels.  For the 0\%--10\% (60\%--70\%) data, 
panels (a) and (c) ((e) and (g)) show pairs reconstructed in the backward 
(-2.2$<$$y$$<$-1.2) and panels (b) and (d) ((f) and (h)) forward 
(1.2$<$$y$$<$2.2) muons arms respectively.  The solid line represents a 
fit to the data using a double Gaussian line shape plus an exponential 
background, see text for details. }
\end{figure*}

The PHENIX detector recorded \cuau events at $\sqrt{s_{NN}}$\,=\,200\,GeV 
during the 2012 data-taking period at the Relativistic Heavy-Ion Collider 
(RHIC) at Brookhaven National Laboratory. The detector is shown 
schematically in Fig.~\ref{fig:Detector}. Global event information is 
obtained from the beam-beam counters (BBC), which comprise two arrays of 
64 quartz \v{C}erenkov counters that measure charged particles within the 
pseudorapidity range ($3.0$$<$$|\eta|$$<$$3.9$). The BBC provides the 
primary level-1 trigger for \cuau minimum bias events, requiring two or 
more hits on each side of the interaction point and a fast reconstructed 
event vertex located along the beam direction within $\pm$30\,cm of the 
nominal center of the PHENIX acceptance. For this analysis, 20.7 billion 
($\mathcal{L}$\,=\,4.3\,nb$^{-1}$) sampled minimum bias events were used 
within $\pm$30\,cm. The corresponding N+N integrated luminosity used is 
53\,pb$^{-1}$.

For the data set used in this analysis the primary level-1 trigger from 
the BBC is required to be in coincidence with an additional level-1 
trigger, requiring two muon candidates to penetrate fully through the muon 
identifier. The trigger logic for a muon candidate requires a road of 
fired Iarocci tubes in at least four planes, including the most downstream 
plane relative to the collision point.

Muons at forward rapidities are reconstructed in this analysis using the 
South and North (see Fig.~\ref{fig:Detector}) muon spectrometers. The muon 
spectrometers comprise four sub-components: a steel absorber, a magnet 
(one per spectrometer), a muon tracker (MuTr), and a muon identifier 
(MuID).  A detailed description of the muon detectors is given 
in~\cite{Akikawa:2005}.  In 2010, an additional 36.2\,cm of steel 
absorbers ($\lambda_{I}$\,=\,2.3) were added to help increase the relative 
yield of muons compared to hadronic background. This additional material 
decreases the efficiency of the low-$p_{T}$ muons which punch through all 
muon arm materials by $\sim$30\%--40\%.  The minimum momentum for a muon 
to reach the outermost MuID plane is 3~GeV/$c$. Three sets of cathode 
strip chambers (MuTr), inside the muon magnet, follow the absorber 
material which are used to measure the momentum of tracks within the 
detector volume.  The final component (MuID) comprises alternating steel 
absorbers and Iarocci tubes, which further reduce the number of hadronic 
tracks which punch through the initial layers of absorber material and 
masquerade as muons.

\section{Data Analysis}

\subsection{Centrality Determination}

The events are sorted into centrality classes using the combined charge 
from both BBC counters. The number of participating nucleons (\Npart) and 
number of binary collisions (\Ncoll) in each centrality class is obtained 
from a Monte Carlo Glauber calculation~\cite{Miller:2007ri} folded with a 
Negative Binomial Distribution that is fitted to the measured BBC charge 
distribution in the charge range where the BBC trigger is fully efficient. 
For peripheral events where the minimum bias trigger is not fully 
efficient, the efficiency is obtained from a comparison of the measured 
BBC charge distribution to the Negative Binomial Distribution. The minimum 
bias trigger is determined to fire on 93\%\,$\pm$\,3\% of the inelastic 
\cuau cross section.

Several baseline parameters are used to characterize the Glauber model 
nuclei and their interactions.  Nucleons in each gold and copper nucleus 
are distributed using a Woods-Saxon function, given in 
Eq.~\ref{eqn:WoodsSaxon}, with a radius, $R$, of 6.38\,fm (Au) and 
4.20\,fm (Cu) along with diffuseness, $a$, of 0.535\,fm and 0.596\,fm 
respectively. A minimum internucleon distance is enforced to be 0.4\,fm 
(known as the hard-core radius) such that nucleons cannot overlap in the 
nucleus.  The nucleon-nucleon inelastic scattering cross section of 42\,mb 
is used as default.

\begin{equation}
\label{eqn:WoodsSaxon}
\ensuremath{\rho(r) = \frac{\rho_{0}}{1+e^{-(R-r)/a}}}
\end{equation} 

The systematic uncertainties on \Npart and \Ncoll are estimated by
varying the baseline parameters to the Glauber model from four sources: 
\begin{enumerate}
\item The nucleon-nucleon inelastic scattering cross section of 42\,mb 
is varied by $\pm$3\,mb.
\item Extreme radii and diffuseness cases were compared to the default 
baseline using (a) $R_{\rm Au}$\,=\,6.25\,fm, 
$a_{\rm Au}$\,=\,0.530\,fm and $R_{\rm Cu}$\,=\,4.11\,fm, 
$a_{\rm Cu}$\,=\,0.590\,fm, and (b) $R_{\rm Au}$\,=\,6.65\,fm, 
$a_{\rm Au}$\,=\,0.550\,fm and $R_{\rm Cu}$\,=\,4.38\,fm, 
$a_{\rm Cu}$\,=\,0.613\,fm.
\item The condition of a minimum internucleon distance was removed such 
that nucleons are allowed to overlap in the initial nucleon distribution.
\item Since the trigger efficiency is 93\% with an uncertainty of 3\%, the 
Glauber parameters are also calculated assuming an efficiency of 90\% and 
96\%.
\end{enumerate}
A total of eight variations (including the baseline) of the Glauber model
conditions are used to estimate the systematic uncertainties.
The extracted total cross section from this Glauber model for Cu+Au
collisions is estimated to be $\sigma_{\rm Cu+Au}$\,=\,5.23\,$\pm$\,0.15\,$b$.
The results are summarized in Table~\ref{tbl:Glauber}.

%%%%%%%%%%%%%%%%%%%%%%%%%%%%%%%%%%%%%%%%%%%%%% Table I
\begin{table}[htb]
  \caption{\label{tbl:Glauber} Glauber-estimated centrality parameters
    in Cu+Au collisions.}
  \begin{ruledtabular} \begin{tabular}{ccccc}
  Centrality & \Ncoll & \Npart & $N_{\rm part}^{Au}$ & $N_{\rm part}^{Cu}$ \\ \hline
   0\%--10\%     & 373.3\,$\pm$\,34.6 & 177.2\,$\pm$\,5.2  &  117.5\,$\pm$\,3.4 &  59.7\,$\pm$\,1.8 \\
  10\%--20\%     & 254.2\,$\pm$\,21.7 & 132.4\,$\pm$\,3.7  &   82.1\,$\pm$\,2.3 &  50.2\,$\pm$\,1.4 \\
  20\%--30\%     & 161.5\,$\pm$\,14.8 &  95.1\,$\pm$\,3.2  &   56.8\,$\pm$\,1.9 &  38.3\,$\pm$\,1.3 \\
  30\%--40\%     &  97.1\,$\pm$\,10.1 &  65.7\,$\pm$\,3.4  &   38.3\,$\pm$\,2.0 &  27.5\,$\pm$\,1.4 \\
  40\%--50\%     &  55.0\,$\pm$\,6.3  &  43.3\,$\pm$\,3.0  &   24.8\,$\pm$\,1.7 &  18.5\,$\pm$\,1.3 \\
  50\%--60\%     &  29.0\,$\pm$\,3.9  &  26.8\,$\pm$\,2.6  &   15.1\,$\pm$\,1.5 &  11.7\,$\pm$\,1.2 \\
  60\%--70\%     &  14.0\,$\pm$\,2.4  &  15.2\,$\pm$\,2.0  &    8.5\,$\pm$\,1.1 &   6.8\,$\pm$\,0.9 \\
  70\%--80\%     &   6.2\,$\pm$\,1.4  &   7.9\,$\pm$\,1.5  &    4.3\,$\pm$\,0.8 &   3.5\,$\pm$\,0.7 \\
  80\%--90\%     &   2.4\,$\pm$\,0.7  &   3.6\,$\pm$\,0.8  &    1.9\,$\pm$\,0.4 &   1.7\,$\pm$\,0.4 \\
  \end{tabular} \end{ruledtabular}
\end{table}

\subsection{Muon-Track Reconstruction}

The data reported here were obtained from the PHENIX muon spectrometers, 
which cover the rapidity ranges \ybackward and \yforward. Muon candidates 
are reconstructed by finding tracks that penetrate through all layers of 
the MuID, then matching these to tracks in the MuTr. The requirement of 
the track penetrating the full absorber material through the MuID 
significantly reduces the hadron contribution. However, with small 
probability (of order $\sim$1/1000) a charged hadron may penetrate the 
material without suffering a hadronic interaction. 
Additionally, the muon spectrometer cannot reject 
most muons that originate from charged pions and kaons which decay before 
the absorber in front of the MuTr. For the dimuon reconstruction in this 
analysis, pairs of muon candidate tracks are selected and a combined fit 
is performed with the collision $z$-vertex from the BBC. We apply various 
cuts to enhance the sample of good muon track pairs, including cuts on the 
individual track $\chi^2$ values, the matching between position and 
direction vectors of the MuID track and the MuTr track projected to the 
front of the MuID, and finally the $\chi^2$ of the track pair and BBC 
$z$-vertex combined fit.

\subsection{$\mu^{+}$+$\mu^{-}$ Analysis}

All opposite charge-sign pairs within an event are combined to form an 
effective invariant mass, see Fig.~\ref{fig:RawCounts}.  Punch-through 
hadrons or single muons can randomly combine to form a combinatorial 
background.  Muon pairs from decays of heavy vector-mesons, the $\psi$ and 
$\Upsilon$ families, form peaks in the mass spectrum. There are continuum 
contributions from correlated muon pairs due to the Drell-Yan process, and 
due to correlated semileptonic open heavy flavor decays.  Owing to the 
momentum resolution in the MuTr, distinct \jpsi and \psip peaks are not 
visible in this analysis.  The left and right panels represent data in the 
most central event class (0\%--10\%) and a mid-peripheral (60\%--70\%) class 
respectively.

The total combinatorial background is estimated using a mixed event 
technique, where oppositely charged tracks from different events are 
combined to form an effective mass (see~\cite{Adare:2011yf} for details).  
As these are independent events, all real correlations are necessarily 
absent and only the combinatorial background remains (open symbols on the 
upper panels of Fig.~\ref{fig:RawCounts}). To extract the yield, a fit is 
made which includes the combinatorial background (from above) plus an 
acceptance-modified~\cite{Adare:2011vq} double-Gaussian line shape which 
represents the \jpsi signal, along with an acceptance-modified exponential 
term to account for the remaining correlated physical background.  The 
double-Gaussian line shape is inspired by the line shape measured in \pp 
collisions~\cite{Adare:2007gn}, only the yield and the \jpsi mass width 
are allowed to vary, the latter accounts for its degradation in the large 
background of heavy-ion collisions. The resultant mass width is found to 
vary linearly with multiplicity in the spectrometer arms from 
0.15\,GeV/$c^{2}$ at low multiplicity to 0.18\,GeV/$c^{2}$ at the highest 
multiplicity in Cu+Au collisions. The fit range is from 1.75 to 5.0 
GeV/$c^2$, and the resultant fit function is shown as a solid line on 
Fig~\ref{fig:RawCounts}.  Systematic uncertainties of 2.2\%-10.6\% (see 
Table~\ref{tbl:Uncertainties}) are associated with the yield extraction to 
account for uncertainty in the combinatorial background subtraction and 
the fit function and fit range used.  Additionally, the extracted yields 
were systematically checked for consistency by using a like-sign 
combinatorial background subtraction method, and good agreement was found.  
A total of 35k \jpsi are counted across all centrality and rapidities.

\subsection{Efficiency and Corrections}

The efficiency for reconstructing the \jpsi in the muon arms is estimated 
by embedding {\sc pythia~5.428}~\cite{Sjostrand:pythia} 
\jpsi$\rightarrow\mumu$ into real minimum bias events (i.e. a sample of 
events which do not necessarily contain a \jpsi candidate). First, the 
{\sc pythia} \jpsi$\rightarrow\mumu$ events are simulated through a full 
{{\sc geant} 3.21}~\cite{Brun:geant} description of the PHENIX detector.  
This simulation accounts for inefficiencies due to dead materials, 
including those due to the additional steel absorber.  The resultant 
simulated hits in the muon tracker and identifier are added to the signals 
found in the real data event.  Once embedded, the amalgamated event is 
passed through the same full reconstruction chain as used for real data.  
The simulations include a trigger emulator.  In the final step, the yield 
of reconstructed \jpsi divided by the originally simulated number of {\sc 
pythia} \jpsi$\rightarrow\mumu$, in the same rapidity range, determines 
the acceptance$\times$efficiency correction factor ($A\epsilon$ in 
Eq.~\ref{eqn:InvYield}).  Depending on which muon spectrometer and the 
centrality, the acceptance$\times$efficiency varies from 2.5\% (3.6\%) 
(0\%--10\% central at positive (negative) rapidity) to 3.4\% (5.2\%) (70\%--80\% 
peripheral).

Uncertainties due to the assumed input {\sc pythia} rapidity and momentum 
distributions for the \jpsi$\rightarrow\mumu$ were previously evaluated 
for the correction factors and were found to be 
$\sim$4\%~\cite{Adare:2012wf}. An efficiency uncertainty of $\sim$10\% 
represents an overall uncertainty on extracting the reconstruction and 
trigger efficiency from the embedding procedure. Small run-to-run 
variations in the detector acceptance and MuID efficiencies were also 
evaluated to be 5\% and 2.8\%, respectively.  These systematic 
uncertainties are added in quadrature for the total uncertainty on the 
measured yields.  An error representing the uncertainty in determining the 
efficiency (10\%) is also added in quadrature to the Type-B systematic 
uncertainty.

%%%%%%%%%%%%%%%%%%%%%%%%%%%%%%%%%%%%%%%%%%%%%% Table II
\begin{table}[htb]
  \caption{\label{tbl:Uncertainties} Estimated systematic uncertainties. }
  \begin{ruledtabular} \begin{tabular}{ccc}
  Source and uncertainty (\%) & Type \\
  \hline
  \jpsi Signal extraction   & $\pm$2.2--10.6 & A \\%stat-subtracted
  run-to-run efficiency variation & $\pm$2.8       & B \\%mutr and miud eff
  Input \jpsi \pt distributions & $\pm$4.0       & B \\%pythia
  Detector acceptance & $\pm$5.0       & B \\%
  Reconstruction and trigger efficiency & $\pm$10.0      & B \\%stat error on efficiency
  Glauber (\Ncoll)    & $\pm$10--29    & B \\%Ncoll error
  \pp reference    & $\pm$7.1    & C \\%pp error
  \end{tabular} \end{ruledtabular}
\end{table}

%%%%%%%%%%%%%%%%%%%%%%%%%%%%%%%%%%%%%%%%%%%%%% Table III
\begin{table}
\caption{\label{tbl:invyields} Invariant yield at forward 
(1.2$<$$y$$<$2.2) and backward (-2.2$<$$y$$<$-1.2) rapidity as a function 
of centrality.  The first and second uncertainties listed represent Type-A 
and Type-B uncertainties, respectively (see text for definitions).  No 
Type-C (global) systematic is assigned.}
\begin{ruledtabular} \begin{tabular}{ccc}
  \multirow{4}{*}{Centrality} 
  & \multicolumn{2}{c}{$B \frac{dN}{dy}\times10^{-6}$} \\
  & Forward & Backward    \\
  & Cu-going direction & Au-going direction   \\
                              & 1.2$<$$y$$<$2.2 & -2.2$<$$y$$<$-1.2    \\ \hline
0\%--10\%   & 60.53\,$\pm$\,6.39\,$\pm$\,7.39 & 68.76\,$\pm$\,3.16\,$\pm$\,8.39 \\
10\%--20\%  & 46.99\,$\pm$\,4.53\,$\pm$\,5.74 & 60.12\,$\pm$\,2.56\,$\pm$\,7.34 \\
20\%--30\%  & 31.50\,$\pm$\,2.80\,$\pm$\,3.85 & 43.31\,$\pm$\,2.97\,$\pm$\,5.29 \\
30\%--40\%  & 22.05\,$\pm$\,1.28\,$\pm$\,2.69 & 29.25\,$\pm$\,1.28\,$\pm$\,3.57 \\
40\%--50\%  & 16.45\,$\pm$\,0.94\,$\pm$\,2.01 & 19.96\,$\pm$\,0.95\,$\pm$\,2.44 \\
50\%--60\%  & 9.92\,$\pm$\,0.57\,$\pm$\,1.21 & 11.95\,$\pm$\,0.80\,$\pm$\,1.46 \\
60\%--70\%  & 5.76\,$\pm$\,0.40\,$\pm$\,0.70 & 6.80\,$\pm$\,0.32\,$\pm$\,0.83 \\
70\%--80\%  & 3.52\,$\pm$\,0.28\,$\pm$\,0.43 & 3.68\,$\pm$\,0.30\,$\pm$\,0.45 \\
80\%--90\%  & 1.44\,$\pm$\,0.20\,$\pm$\,0.18 & 1.59\,$\pm$\,0.14\,$\pm$\,0.19 \\
  \end{tabular} \end{ruledtabular}
\end{table}

%%THIS DOES INCLUDE NCOLL ERROR IN TYPE-B
%%%%%%%%%%%%%%%%%%%%%%%%%%%%%%%%%%%%%%%%%%%%%% Table IV
\begin{table*}[thb]
\caption{\label{tbl:Results} 
Nuclear modification factor ($\raa$) at forward (1.2$<$$y$$<$2.2 -- 
Cu-going) and backward (-2.2$<$$y$$<$-1.2 -- Au-going) rapidity and 
forward/backward ratio as a function of centrality.  The first and second 
uncertainties listed represent Type-A and Type-B uncertainties, 
respectively (see text for definitions).  An additional 7.1\% Type-C 
(global) systematic also applies for the $\raa$.}
  \begin{ruledtabular} \begin{tabular}{cccccc}
  \multirow{4}{*}{Centrality} & \multicolumn{2}{c}{$\raa$}              &  \\
                              & Forward & Backward                        & Forward/Backward \\
                              & Cu-going direction & Au-going direction   & Ratio \\
                              & 1.2$<$$y$$<$2.2 & -2.2$<$$y$$<$-1.2       & \\ \hline
0\%--10\%  & 0.239 $\pm$ 0.025 $\pm$ 0.037 & 0.271 $\pm$ 0.012 $\pm$ 0.042  & 0.88 $\pm$ 0.10 $\pm$ 0.14 \\
10\%--20\%  & 0.272 $\pm$ 0.026 $\pm$ 0.040 & 0.348 $\pm$ 0.015 $\pm$ 0.052 & 0.78 $\pm$ 0.08 $\pm$ 0.13 \\
20\%--30\%  & 0.287 $\pm$ 0.026 $\pm$ 0.044 & 0.394 $\pm$ 0.027 $\pm$ 0.060 & 0.73 $\pm$ 0.08 $\pm$ 0.12 \\
30\%--40\%  & 0.334 $\pm$ 0.019 $\pm$ 0.054 & 0.443 $\pm$ 0.019 $\pm$ 0.071 & 0.75 $\pm$ 0.05 $\pm$ 0.12 \\
40\%--50\%  & 0.440 $\pm$ 0.025 $\pm$ 0.074 & 0.534 $\pm$ 0.025 $\pm$ 0.089 & 0.82 $\pm$ 0.06 $\pm$ 0.13 \\
50\%--60\%  & 0.486 $\pm$ 0.028 $\pm$ 0.087 & 0.586 $\pm$ 0.039 $\pm$ 0.104 & 0.83 $\pm$ 0.07 $\pm$ 0.14 \\
60\%--70\%  & 0.605 $\pm$ 0.042 $\pm$ 0.127 & 0.714 $\pm$ 0.034 $\pm$ 0.150 & 0.85 $\pm$ 0.07 $\pm$ 0.14 \\
70\%--80\%  & 0.835 $\pm$ 0.065 $\pm$ 0.214 & 0.873 $\pm$ 0.072 $\pm$ 0.224 & 0.96 $\pm$ 0.11 $\pm$ 0.16 \\
80\%--90\%  & 0.875 $\pm$ 0.124 $\pm$ 0.268 & 0.968 $\pm$ 0.084 $\pm$ 0.296 & 0.90 $\pm$ 0.15 $\pm$ 0.15 \\
  \end{tabular} \end{ruledtabular}
\end{table*}

The invariant \jpsi yields ($\frac{dN}{dy}$) are calculated for the 
\jpsi$\rightarrow\mumu$ branching fraction, $B$, from

\begin{equation}
\label{eqn:InvYield}
\ensuremath{B \frac{dN}{dy} = \frac{1}{N_{\rm event}} \frac{N_{\rm measured}^{J\psi}}{\Delta yA\epsilon}}
\end{equation} 

\noindent $N^{J/\psi}_{\rm measured}$ is the number of measured \jpsi per 
unit rapidity ($\Delta y$).  The number of minimum-bias equivalent events is 
given by $N_{\rm event}$.

\section{Results}

The invariant yields calculated using Eq.~\ref{eqn:InvYield} are 
summarized in Table~\ref{tbl:invyields}. The nuclear modification factor, 
$\raa$ is formed from the invariant yields using Eq.~\ref{eqn:RAA},

\begin{equation}
\label{eqn:RAA}
\ensuremath{R_{AA} = \frac{1}{\langle \Ncoll 
\rangle}\frac{dN({\rm CuAu})/dy}{dN(pp)/dy}},
\end{equation} 

\noindent where $dN({\rm CuAu})/dy$ and $dN(pp)/dy$ represent the 
invariant yields measured in Cu+Au and \pp collisions, respectively. Data 
from the same detector recorded in 2006 and 2008 are used as the reference 
\pp data~\cite{Adare:2010fn}.

%%%%%%%%%%%%%%%%%%%%%%%%%%%%%%%%%%%%%%%%%%%%%%%%%%%%%%%%%% Fig_3
\begin{figure*}[htb]
  \begin{minipage}{0.64\linewidth}
    \includegraphics[width=1.0\linewidth]{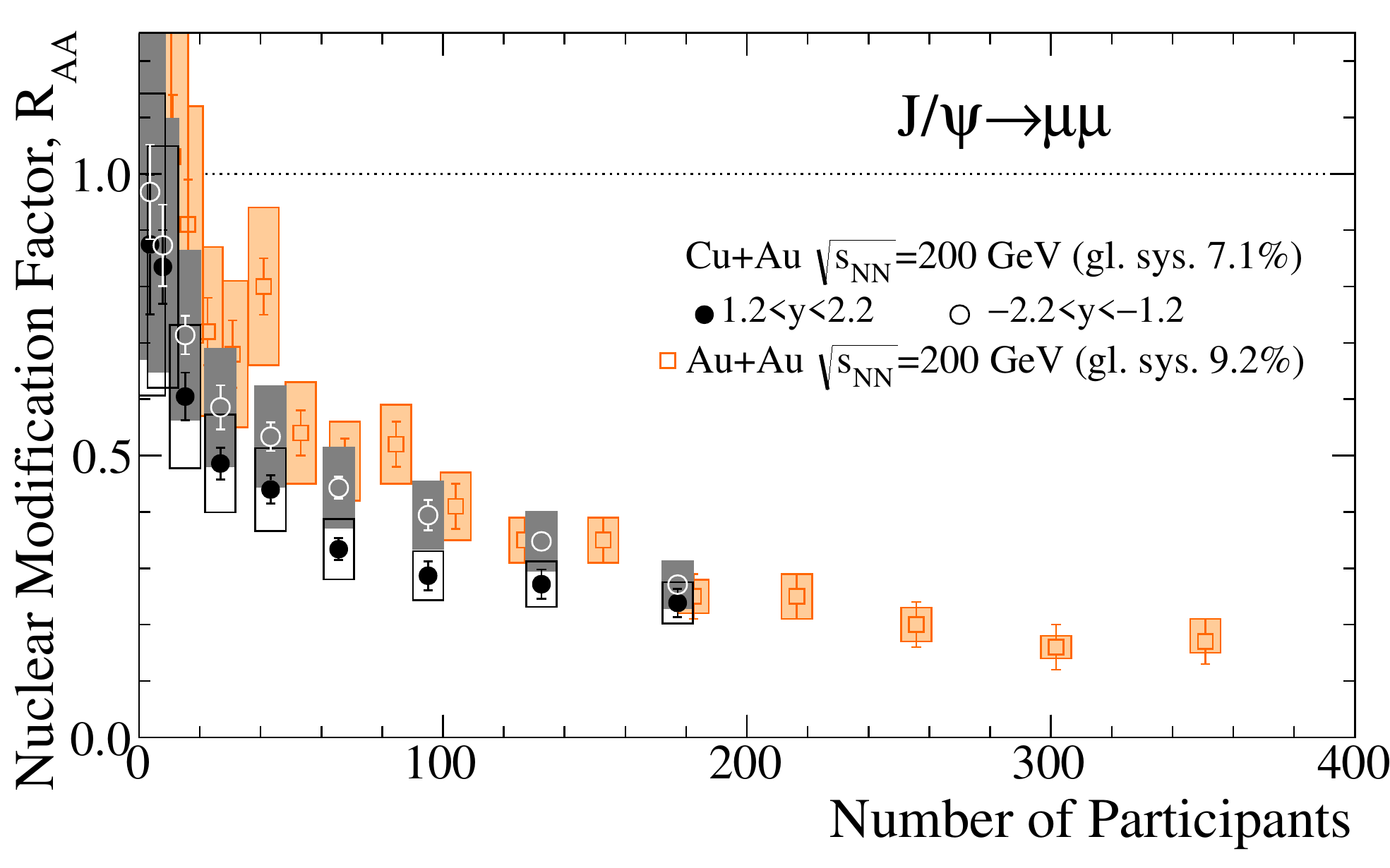}
  \end{minipage}
  \hspace{0.025\linewidth}
  \begin{minipage}{0.29\linewidth}
    \caption{\label{fig:RAA}(Color online) 
Nuclear modification factor, \raa, measured as a
function of collision centrality ($\Npart$).  Values for  $\jpsi$
at forward (Cu-going) rapidity are shown as closed circles and at
backward (Au-going) rapidity as open circles. For reference, Au$+$Au
data~\cite{Adare:2011yf} are also shown, averaged over forward and
backward rapidities, as red squares.
}
  \end{minipage}
\end{figure*}

%%%%%%%%%%%%%%%%%%%%%%%%%%%%%%%%%%%%%%%%%%%%%%% Fig_4
\begin{figure*}[htb]
  \begin{minipage}{0.64\linewidth}
    \includegraphics[width=1.0\linewidth]{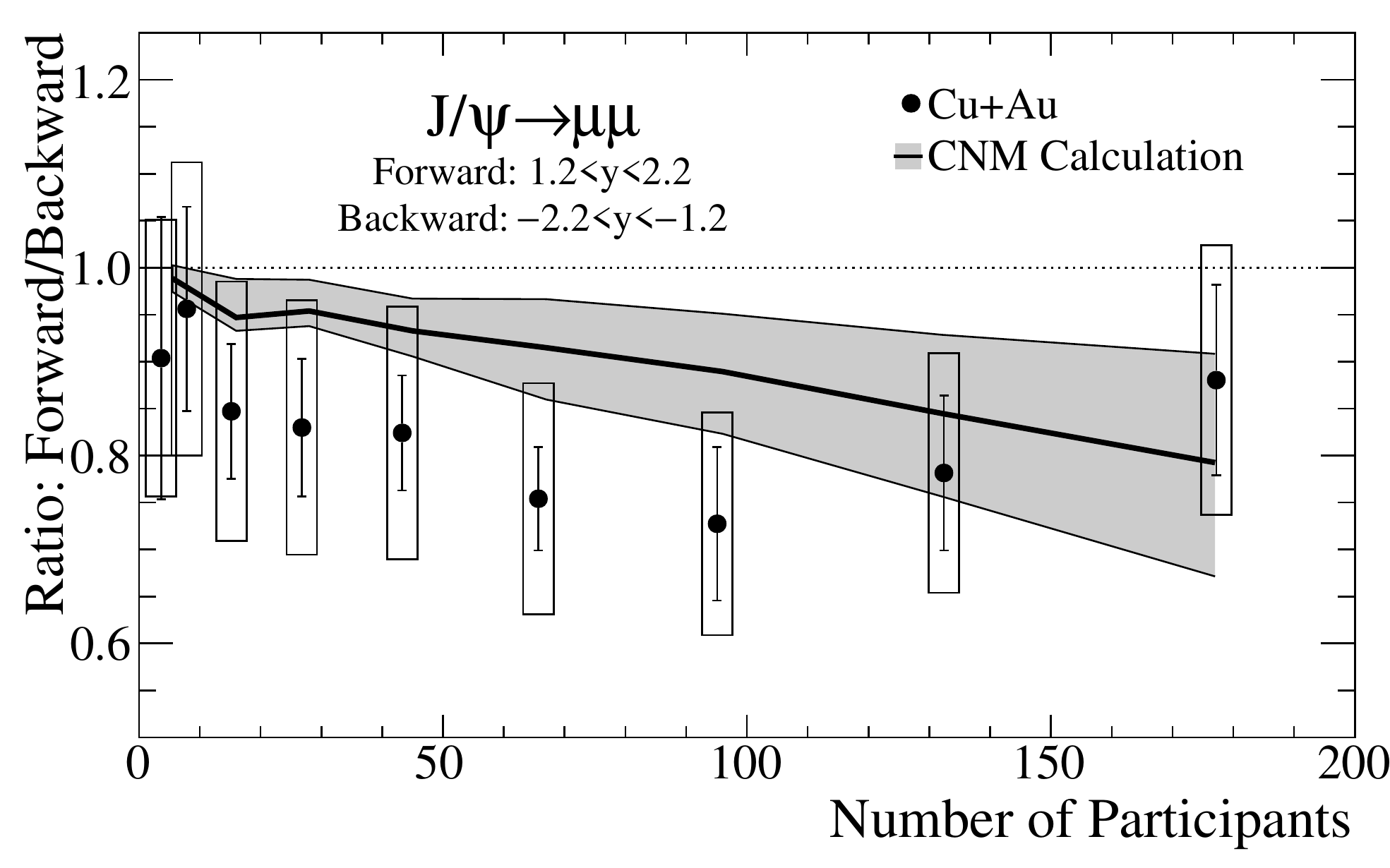}
  \end{minipage}
  \hspace{0.025\linewidth}
  \begin{minipage}{0.29\linewidth}
    \caption{\label{fig:Ratio}
      Ratio of forward- to backward-rapidity (Cu-going/Au-going) \jpsi yields measured in Cu+Au
      collisions (symbols).  Also shown
      is a model~\cite{Nagle:2010ix} which estimates the contribution
      from cold nuclear matter; the band represents the extreme
      nPDF parameter sets as described in \cite{Eskola:2009uj}.
    }
  \end{minipage}
\end{figure*}

The values of \raa versus centrality are listed in Table~\ref{tbl:Results} 
and shown as a function of \Npart in Fig.~\ref{fig:RAA}. The \raa for 
Au$+$Au collisions~\cite{Adare:2011yf} at the same collision energy and 
rapidity (red squares) is shown in Fig.~\ref{fig:RAA} for comparison. The 
dependence of the \cuau nuclear modification on \Npart at backward 
(Au-going) rapidity is similar to that for Au$+$Au collisions, while the 
\cuau \raa at forward (Cu-going) rapidity is noticeably lower.

The uncertainties on the measured yield values are separated into three 
types.  Type-A uncertainties are random point-to-point uncertainties which 
are combined in quadrature with the statistical uncertainty associated 
with each data point.  These are represented by vertical bars in the 
figures.  Type-B uncertainties are correlated point-to-point systematic 
uncertainties which are represented by boxes in the figures.  Type-C 
uncertainties represent a global systematic scale uncertainty, which 
represents the scale uncertainty from the measured \pp reference data. The 
values of the point-to-point systematic uncertainties are summarized in 
Table~\ref{tbl:Uncertainties}.

Forward and backward differences can be observed when forming the ratio of 
the yield values for the forward rapidity to the backward rapidity. This 
is shown in Fig.~\ref{fig:Ratio}, and the values are presented in 
Table~\ref{tbl:Results}.  This ratio has the advantage of reduced 
systematic uncertainties due to the cancellation of type-C and some type-B 
correlated uncertainties that apply to \raa, those which are related to 
the Glauber model calculation. The 20\%--30\% difference in suppression 
between forward and backward rapidity \raa evident in Fig.~\ref{fig:Ratio} 
could be due to hot matter effects, CNM effects, or a combination of both.

To obtain an indication of the expected size of the difference due to CNM 
effects, we use a simple Glauber model that combines gluon modifications 
as a function of Bjorken $x$ and $Q^2$, taken from the EPS09 shadowing 
parametrization \cite{Eskola:2009uj}, and a single effective $c\bar{c}$ 
break up cross section (4\,mb) that approximately reproduces the \dau 
nuclear modification observed in PHENIX data across all rapidities~ 
\cite{Nagle:2010ix}. It should be emphasized that this simple model uses a 
constant effective $c\bar{c}$ cross section to account for nonshadowing 
effects at all rapidities, while in fact both breakup and energy loss 
contributions are expected to be rapidity dependent. Thus the calculation 
reflects only the expected difference in shadowing between forward and 
backward rapidity in \cuau. The calculation, shown in Fig.~\ref{fig:Ratio} 
indicates that the size of the expected shadowing difference is comparable 
with the effect seen in the data, and has the same sign.

Hot matter effects are expected to be greater at backward rapidity in 
\cuau collisions, where the particle multiplicity should be about 20\% 
higher in the Au-going direction than in the Cu-going 
direction~\cite{Chen:2005zy}. Increased suppression due to higher energy 
density at backward rapidity would lead to an increase in the ratio shown 
in Fig.~\ref{fig:Ratio}.  Increased recombination effects may also occur 
at higher energy density (see for example~\cite{Zhao:2010nk}), increasing 
the \jpsi yield and tending to decrease the ratio shown in 
Fig.~\ref{fig:Ratio}.

The new rapidity dependent \cuau \jpsi data presented here form part of a 
large \jpsi data set at RHIC energies that includes \pp, \dau, \cucu and 
\auau collision data. These \jpsi nuclear modification data result from a 
varied mix of energy densities and cold nuclear matter effects, providing 
a broad range of conditions with which to confront models of \jpsi 
production.

\section{Summary and Conclusions}

We have measured the centrality dependence of \jpsi production in 
asymmetric Cu+Au collisions.  We find the centrality evolution of the 
nuclear modification (\raa) at backward rapidity to be similar to that 
measured in Au$+$Au collisions at the same number of participants, while at 
forward rapidity (the Cu-going direction) it is significantly smaller. At 
backward rapidity, in the most central 10\% collisions, 
\raa\,=\,0.271\,$\pm$\,0.012\,$\pm$\,0.042. At forward rapidity the 
suppression is on average about 20\% stronger in the centrality range 
0\%--40\%, while for the most peripheral collisions the ratio is consistent 
with unity within systematic uncertainties.

The difference between forward (Cu-going) and backward (Au-going) \jpsi 
modification is found to be comparable in magnitude and of the same sign 
as the expected difference from shadowing effects. These data add a 
completely new admixture of hot and cold nuclear matter effects to those 
already sampled for \jpsi production at RHIC energies, broadening the 
range of conditions with which models of \jpsi production can be 
confronted.

%%%%%%%%%%%%%%%%%%%%%%%%%  Acknowledgements 

\section*{ACKNOWLEDGMENTS}   % 2012 long form for all journals

We thank the staff of the Collider-Accelerator and Physics
Departments at Brookhaven National Laboratory and the staff of
the other PHENIX participating institutions for their vital
contributions.  We acknowledge support from the 
Office of Nuclear Physics in the
Office of Science of the Department of Energy, the
National Science Foundation, Abilene Christian University
Research Council, Research Foundation of SUNY, and Dean of the
College of Arts and Sciences, Vanderbilt University (U.S.A),
Ministry of Education, Culture, Sports, Science, and Technology
and the Japan Society for the Promotion of Science (Japan),
Conselho Nacional de Desenvolvimento Cient\'{\i}fico e
Tecnol{\'o}gico and Funda\c c{\~a}o de Amparo {\`a} Pesquisa do
Estado de S{\~a}o Paulo (Brazil),
Natural Science Foundation of China (P.~R.~China),
Ministry of Science, Education, and Sports (Croatia),
Ministry of Education, Youth and Sports (Czech Republic),
Centre National de la Recherche Scientifique, Commissariat
{\`a} l'{\'E}nergie Atomique, and Institut National de Physique
Nucl{\'e}aire et de Physique des Particules (France),
Bundesministerium f\"ur Bildung und Forschung, Deutscher
Akademischer Austausch Dienst, and Alexander von Humboldt Stiftung (Germany),
Hungarian National Science Fund, OTKA (Hungary), 
Department of Atomic Energy and Department of Science and Technology (India), 
Israel Science Foundation (Israel), 
National Research Foundation of Korea of the Ministry of Science,
ICT, and Future Planning (Korea),
Physics Department, Lahore University of Management Sciences (Pakistan),
Ministry of Education and Science, Russian Academy of Sciences,
Federal Agency of Atomic Energy (Russia),
VR and Wallenberg Foundation (Sweden), 
the U.S. Civilian Research and Development Foundation for the
Independent States of the Former Soviet Union, 
the Hungarian American Enterprise Scholarship Fund,
and the US-Israel Binational Science Foundation.

%%%%%%%%%%%%%%%%%%%%%%%%%%%  References 

%\bibliography{ppg163x0}

\begin{thebibliography}{45}%
\makeatletter
\providecommand \@ifxundefined [1]{%
 \@ifx{#1\undefined}
}%
\providecommand \@ifnum [1]{%
 \ifnum #1\expandafter \@firstoftwo
 \else \expandafter \@secondoftwo
 \fi
}%
\providecommand \@ifx [1]{%
 \ifx #1\expandafter \@firstoftwo
 \else \expandafter \@secondoftwo
 \fi
}%
\providecommand \natexlab [1]{#1}%
\providecommand \enquote  [1]{``#1''}%
\providecommand \bibnamefont  [1]{#1}%
\providecommand \bibfnamefont [1]{#1}%
\providecommand \citenamefont [1]{#1}%
\providecommand \href@noop [0]{\@secondoftwo}%
\providecommand \href [0]{\begingroup \@sanitize@url \@href}%
\providecommand \@href[1]{\@@startlink{#1}\@@href}%
\providecommand \@@href[1]{\endgroup#1\@@endlink}%
\providecommand \@sanitize@url [0]{\catcode `\\12\catcode `\$12\catcode
  `\&12\catcode `\#12\catcode `\^12\catcode `\_12\catcode `\%12\relax}%
\providecommand \@@startlink[1]{}%
\providecommand \@@endlink[0]{}%
\providecommand \url  [0]{\begingroup\@sanitize@url \@url }%
\providecommand \@url [1]{\endgroup\@href {#1}{\urlprefix }}%
\providecommand \urlprefix  [0]{URL }%
\providecommand \Eprint [0]{\href }%
\providecommand \doibase [0]{http://dx.doi.org/}%
\providecommand \selectlanguage [0]{\@gobble}%
\providecommand \bibinfo  [0]{\@secondoftwo}%
\providecommand \bibfield  [0]{\@secondoftwo}%
\providecommand \translation [1]{[#1]}%
\providecommand \BibitemOpen [0]{}%
\providecommand \bibitemStop [0]{}%
\providecommand \bibitemNoStop [0]{.\EOS\space}%
\providecommand \EOS [0]{\spacefactor3000\relax}%
\providecommand \BibitemShut  [1]{\csname bibitem#1\endcsname}%
\let\auto@bib@innerbib\@empty
%</preamble>
\bibitem [{\citenamefont {Matsui}\ and\ \citenamefont
  {Satz}(1986)}]{Matsui:1986dk}%
  \BibitemOpen
  \bibfield  {author} {\bibinfo {author} {\bibfnamefont {T.}~\bibnamefont
  {Matsui}}\ and\ \bibinfo {author} {\bibfnamefont {H.}~\bibnamefont {Satz}},\
  }\href {\doibase 10.1016/0370-2693(86)91404-8} {\bibfield  {journal}
  {\bibinfo  {journal} {Phys. Lett. B}\ }\textbf {\bibinfo {volume} {178}},\
  \bibinfo {pages} {416} (\bibinfo {year} {1986})}\BibitemShut {NoStop}%
%%CITATION = PHLTA,B178,416;%%
\bibitem [{\citenamefont {Frawley}\ \emph {et~al.}(2008)\citenamefont
  {Frawley}, \citenamefont {Ullrich},\ and\ \citenamefont
  {Vogt}}]{Frawley:2008kk}%
  \BibitemOpen
  \bibfield  {author} {\bibinfo {author} {\bibfnamefont {A.~D.}\ \bibnamefont
  {Frawley}}, \bibinfo {author} {\bibfnamefont {T.}~\bibnamefont {Ullrich}}, \
  and\ \bibinfo {author} {\bibfnamefont {R.}~\bibnamefont {Vogt}},\ }\href
  {\doibase 10.1016/j.physrep.2008.04.002} {\bibfield  {journal} {\bibinfo
  {journal} {Phys. Rept.}\ }\textbf {\bibinfo {volume} {462}},\ \bibinfo
  {pages} {125} (\bibinfo {year} {2008})}\BibitemShut {NoStop}%
\bibitem [{\citenamefont {Brambilla}\ \emph {et~al.}(2011)\citenamefont
  {Brambilla}, \citenamefont {Eidelman}, \citenamefont {Heltsley},
  \citenamefont {Vogt}, \citenamefont {Bodwin} \emph
  {et~al.}}]{Brambilla:2010cs}%
  \BibitemOpen
  \bibfield  {author} {\bibinfo {author} {\bibfnamefont {N.}~\bibnamefont
  {Brambilla}}, \bibinfo {author} {\bibfnamefont {S.}~\bibnamefont {Eidelman}},
  \bibinfo {author} {\bibfnamefont {B.~K.}\ \bibnamefont {Heltsley}}, \bibinfo
  {author} {\bibfnamefont {R.}~\bibnamefont {Vogt}}, \bibinfo {author}
  {\bibfnamefont {G.~T.}\ \bibnamefont {Bodwin}},  \emph {et~al.},\ }\href
  {\doibase 10.1140/epjc/s10052-010-1534-9} {\bibfield  {journal} {\bibinfo
  {journal} {Eur. Phys. J. C}\ }\textbf {\bibinfo {volume} {71}},\ \bibinfo
  {pages} {1534} (\bibinfo {year} {2011})}\BibitemShut {NoStop}%
%%CITATION = ARXIV:1010.5827;%%
\bibitem [{\citenamefont {Alessandro}\ \emph {et~al.}(2005)\citenamefont
  {Alessandro} \emph {et~al.}}]{Alessandro:2004ap}%
  \BibitemOpen
  \bibfield  {author} {\bibinfo {author} {\bibfnamefont {B.}~\bibnamefont
  {Alessandro}} \emph {et~al.} (\bibinfo {collaboration} {NA50
  Collaboration}),\ }\href {\doibase 10.1140/epjc/s2004-02107-9} {\bibfield
  {journal} {\bibinfo  {journal} {Eur. Phys. J. C}\ }\textbf {\bibinfo {volume}
  {39}},\ \bibinfo {pages} {335} (\bibinfo {year} {2005})}\BibitemShut
  {NoStop}%
\bibitem [{\citenamefont {Arnaldi}\ \emph {et~al.}(2007)\citenamefont {Arnaldi}
  \emph {et~al.}}]{Arnaldi:2007zz}%
  \BibitemOpen
  \bibfield  {author} {\bibinfo {author} {\bibfnamefont {R.}~\bibnamefont
  {Arnaldi}} \emph {et~al.} (\bibinfo {collaboration} {NA60 Collaboration}),\
  }\href {\doibase 10.1103/PhysRevLett.99.132302} {\bibfield  {journal}
  {\bibinfo  {journal} {Phys. Rev. Lett.}\ }\textbf {\bibinfo {volume} {99}},\
  \bibinfo {pages} {132302} (\bibinfo {year} {2007})}\BibitemShut {NoStop}%
%%CITATION = PRLTA,99,132302;%%
\bibitem [{\citenamefont {Arnaldi}\ \emph {et~al.}(2012)\citenamefont {Arnaldi}
  \emph {et~al.}}]{Arnaldi:2010ky}%
  \BibitemOpen
  \bibfield  {author} {\bibinfo {author} {\bibfnamefont {R.}~\bibnamefont
  {Arnaldi}} \emph {et~al.} (\bibinfo {collaboration} {NA60 Collaboration}),\
  }\href {\doibase 10.1016/j.physletb.2011.11.042} {\bibfield  {journal}
  {\bibinfo  {journal} {Phys. Lett. B}\ }\textbf {\bibinfo {volume} {706}},\
  \bibinfo {pages} {263} (\bibinfo {year} {2012})}\BibitemShut {NoStop}%
%%CITATION = ARXIV:1004.5523;%%
\bibitem [{\citenamefont {Arnaldi}(2009)}]{Arnaldi:2009ph}%
  \BibitemOpen
  \bibfield  {author} {\bibinfo {author} {\bibfnamefont {R.}~\bibnamefont
  {Arnaldi}} (\bibinfo {collaboration} {NA60 Collaboration}),\ }\href {\doibase
  10.1016/j.nuclphysa.2009.10.030} {\bibfield  {journal} {\bibinfo  {journal}
  {Nucl. Phys. A}\ }\textbf {\bibinfo {volume} {830}},\ \bibinfo {pages} {345C}
  (\bibinfo {year} {2009})}\BibitemShut {NoStop}%
%%CITATION = ARXIV:0907.5004;%%
\bibitem [{\citenamefont {Abreu}\ \emph {et~al.}(1999)\citenamefont {Abreu}
  \emph {et~al.}}]{Abreu:1999nn}%
  \BibitemOpen
  \bibfield  {author} {\bibinfo {author} {\bibfnamefont {M.}~\bibnamefont
  {Abreu}} \emph {et~al.} (\bibinfo {collaboration} {NA50 Collaboration}),\
  }\href {\doibase 10.1016/S0370-2693(99)01108-9} {\bibfield  {journal}
  {\bibinfo  {journal} {Phys. Lett. B}\ }\textbf {\bibinfo {volume} {466}},\
  \bibinfo {pages} {408} (\bibinfo {year} {1999})}\BibitemShut {NoStop}%
%%CITATION = PHLTA,B466,408;%%
\bibitem [{\citenamefont {Adare}\ \emph
  {et~al.}(2012{\natexlab{a}})\citenamefont {Adare} \emph
  {et~al.}}]{Adare:2011vq}%
  \BibitemOpen
  \bibfield  {author} {\bibinfo {author} {\bibfnamefont {A.}~\bibnamefont
  {Adare}} \emph {et~al.} (\bibinfo {collaboration} {PHENIX Collaboration}),\
  }\href {\doibase 10.1103/PhysRevD.85.092004} {\bibfield  {journal} {\bibinfo
  {journal} {Phys. Rev. D}\ }\textbf {\bibinfo {volume} {85}},\ \bibinfo
  {pages} {092004} (\bibinfo {year} {2012}{\natexlab{a}})}\BibitemShut
  {NoStop}%
%%CITATION = ARXIV:1105.1966;%%
\bibitem [{\citenamefont {Adare}\ \emph
  {et~al.}(2011{\natexlab{a}})\citenamefont {Adare} \emph
  {et~al.}}]{Adare:2010fn}%
  \BibitemOpen
  \bibfield  {author} {\bibinfo {author} {\bibfnamefont {A.}~\bibnamefont
  {Adare}} \emph {et~al.} (\bibinfo {collaboration} {PHENIX Collaboration}),\
  }\href {\doibase 10.1103/PhysRevLett.107.142301} {\bibfield  {journal}
  {\bibinfo  {journal} {Phys. Rev. Lett.}\ }\textbf {\bibinfo {volume} {107}},\
  \bibinfo {pages} {142301} (\bibinfo {year} {2011}{\natexlab{a}})}\BibitemShut
  {NoStop}%
%%CITATION = ARXIV:1010.1246;%%
\bibitem [{\citenamefont {Adare}\ \emph
  {et~al.}(2013{\natexlab{a}})\citenamefont {Adare} \emph
  {et~al.}}]{Adare:2012qf}%
  \BibitemOpen
  \bibfield  {author} {\bibinfo {author} {\bibfnamefont {A.}~\bibnamefont
  {Adare}} \emph {et~al.} (\bibinfo {collaboration} {PHENIX Collaboration}),\
  }\href {\doibase 10.1103/PhysRevC.87.034904} {\bibfield  {journal} {\bibinfo
  {journal} {Phys. Rev. C}\ }\textbf {\bibinfo {volume} {87}},\ \bibinfo
  {pages} {034904} (\bibinfo {year} {2013}{\natexlab{a}})}\BibitemShut
  {NoStop}%
\bibitem [{\citenamefont {Adare}\ \emph
  {et~al.}(2008{\natexlab{a}})\citenamefont {Adare} \emph
  {et~al.}}]{Adare:2008sh}%
  \BibitemOpen
  \bibfield  {author} {\bibinfo {author} {\bibfnamefont {A.}~\bibnamefont
  {Adare}} \emph {et~al.} (\bibinfo {collaboration} {PHENIX Collaboration}),\
  }\href {\doibase 10.1103/PhysRevLett.101.122301} {\bibfield  {journal}
  {\bibinfo  {journal} {Phys. Rev. Lett.}\ }\textbf {\bibinfo {volume} {101}},\
  \bibinfo {pages} {122301} (\bibinfo {year} {2008}{\natexlab{a}})}\BibitemShut
  {NoStop}%
\bibitem [{\citenamefont {Adare}\ \emph
  {et~al.}(2011{\natexlab{b}})\citenamefont {Adare} \emph
  {et~al.}}]{Adare:2011yf}%
  \BibitemOpen
  \bibfield  {author} {\bibinfo {author} {\bibfnamefont {A.}~\bibnamefont
  {Adare}} \emph {et~al.} (\bibinfo {collaboration} {PHENIX Collaboration}),\
  }\href {\doibase 10.1103/PhysRevC.84.054912} {\bibfield  {journal} {\bibinfo
  {journal} {Phys. Rev. C}\ }\textbf {\bibinfo {volume} {84}},\ \bibinfo
  {pages} {054912} (\bibinfo {year} {2011}{\natexlab{b}})}\BibitemShut
  {NoStop}%
\bibitem [{\citenamefont {Adare}\ \emph {et~al.}(2007)\citenamefont {Adare}
  \emph {et~al.}}]{Adare:2006ns}%
  \BibitemOpen
  \bibfield  {author} {\bibinfo {author} {\bibfnamefont {A.}~\bibnamefont
  {Adare}} \emph {et~al.} (\bibinfo {collaboration} {PHENIX Collaboration}),\
  }\href@noop {} {\bibfield  {journal} {\bibinfo  {journal} {Phys. Rev. Lett.}\
  }\textbf {\bibinfo {volume} {98}},\ \bibinfo {pages} {232301} (\bibinfo
  {year} {2007})}\BibitemShut {NoStop}%
%%CITATION = NUCL-EX/0611020;%%
\bibitem [{\citenamefont {Abelev}\ \emph {et~al.}(2012)\citenamefont {Abelev}
  \emph {et~al.}}]{Abelev:2012kr}%
  \BibitemOpen
  \bibfield  {author} {\bibinfo {author} {\bibfnamefont {B.}~\bibnamefont
  {Abelev}} \emph {et~al.} (\bibinfo {collaboration} {ALICE Collaboration}),\
  }\href {\doibase 10.1016/j.physletb.2012.10.078} {\bibfield  {journal}
  {\bibinfo  {journal} {Phys. Lett. B}\ }\textbf {\bibinfo {volume} {718}},\
  \bibinfo {pages} {295} (\bibinfo {year} {2012})}\BibitemShut {NoStop}%
\bibitem [{\citenamefont {Aamodt}\ \emph {et~al.}(2011)\citenamefont {Aamodt}
  \emph {et~al.}}]{Aamodt:2011gj}%
  \BibitemOpen
  \bibfield  {author} {\bibinfo {author} {\bibfnamefont {K.}~\bibnamefont
  {Aamodt}} \emph {et~al.} (\bibinfo {collaboration} {ALICE Collaboration}),\
  }\href {\doibase 10.1016/j.physletb.2011.09.054,
  10.1016/j.physletb.2012.10.060} {\bibfield  {journal} {\bibinfo  {journal}
  {Phys. Lett. B}\ }\textbf {\bibinfo {volume} {704}},\ \bibinfo {pages} {442}
  (\bibinfo {year} {2011})}\BibitemShut {NoStop}%
\bibitem [{\citenamefont {Abelev}\ \emph {et~al.}({\natexlab{a}})\citenamefont
  {Abelev} \emph {et~al.}}]{Abelev:2013yxa}%
  \BibitemOpen
  \bibfield  {author} {\bibinfo {author} {\bibfnamefont {B.~B.}\ \bibnamefont
  {Abelev}} \emph {et~al.} (\bibinfo {collaboration} {ALICE Collaboration}),\
  }\href@noop {} {} ({\natexlab{a}}),\ \bibinfo {note}
  {arXiv:1308.6726}\BibitemShut {NoStop}%
\bibitem [{\citenamefont {Abelev}\ \emph {et~al.}({\natexlab{b}})\citenamefont
  {Abelev} \emph {et~al.}}]{Abelev:2013ila}%
  \BibitemOpen
  \bibfield  {author} {\bibinfo {author} {\bibfnamefont {B.~B.}\ \bibnamefont
  {Abelev}} \emph {et~al.} (\bibinfo {collaboration} {ALICE Collaboration}),\
  }\href@noop {} {} ({\natexlab{b}}),\ \bibinfo {note}
  {arXiv:1311.0214}\BibitemShut {NoStop}%
\bibitem [{\citenamefont {Leitch}\ \emph {et~al.}(2000)\citenamefont {Leitch}
  \emph {et~al.}}]{Leitch:1999ea}%
  \BibitemOpen
  \bibfield  {author} {\bibinfo {author} {\bibfnamefont {M.}~\bibnamefont
  {Leitch}} \emph {et~al.} (\bibinfo {collaboration} {FNAL E866/NuSea
  Collaboration}),\ }\href {\doibase 10.1103/PhysRevLett.84.3256} {\bibfield
  {journal} {\bibinfo  {journal} {Phys. Rev. Lett.}\ }\textbf {\bibinfo
  {volume} {84}},\ \bibinfo {pages} {3256} (\bibinfo {year}
  {2000})}\BibitemShut {NoStop}%
%%CITATION = NUCL-EX/9909007;%%
\bibitem [{\citenamefont {Abt}\ \emph {et~al.}(2009)\citenamefont {Abt} \emph
  {et~al.}}]{Abt:2008ya}%
  \BibitemOpen
  \bibfield  {author} {\bibinfo {author} {\bibfnamefont {I.}~\bibnamefont
  {Abt}} \emph {et~al.} (\bibinfo {collaboration} {HERA-B Collaboration}),\
  }\href {\doibase 10.1140/epjc/s10052-009-0965-7} {\bibfield  {journal}
  {\bibinfo  {journal} {Eur. Phys. J. C}\ }\textbf {\bibinfo {volume} {60}},\
  \bibinfo {pages} {525} (\bibinfo {year} {2009})}\BibitemShut {NoStop}%
%%CITATION = ARXIV:0812.0734;%%
\bibitem [{\citenamefont {Badier}\ \emph {et~al.}(1983)\citenamefont {Badier}
  \emph {et~al.}}]{Badier:1983dg}%
  \BibitemOpen
  \bibfield  {author} {\bibinfo {author} {\bibfnamefont {J.}~\bibnamefont
  {Badier}} \emph {et~al.} (\bibinfo {collaboration} {NA3 Collaboration}),\
  }\href {\doibase 10.1007/BF01573213} {\bibfield  {journal} {\bibinfo
  {journal} {Z. Phys. C}\ }\textbf {\bibinfo {volume} {20}},\ \bibinfo {pages}
  {101} (\bibinfo {year} {1983})}\BibitemShut {NoStop}%
%%CITATION = ZEPYA,C20,101;%%
\bibitem [{\citenamefont {Alessandro}\ \emph {et~al.}(2004)\citenamefont
  {Alessandro} \emph {et~al.}}]{Alessandro:2003pc}%
  \BibitemOpen
  \bibfield  {author} {\bibinfo {author} {\bibfnamefont {B.}~\bibnamefont
  {Alessandro}} \emph {et~al.} (\bibinfo {collaboration} {NA50
  Collaboration}),\ }\href {\doibase 10.1140/epjc/s2003-01539-y} {\bibfield
  {journal} {\bibinfo  {journal} {Eur. Phys. J. C}\ }\textbf {\bibinfo {volume}
  {33}},\ \bibinfo {pages} {31} (\bibinfo {year} {2004})}\BibitemShut {NoStop}%
%%CITATION = EPHJA,C33,31;%%
\bibitem [{\citenamefont {Alessandro}\ \emph {et~al.}(2006)\citenamefont
  {Alessandro} \emph {et~al.}}]{Alessandro:2006jt}%
  \BibitemOpen
  \bibfield  {author} {\bibinfo {author} {\bibfnamefont {B.}~\bibnamefont
  {Alessandro}} \emph {et~al.} (\bibinfo {collaboration} {NA50
  Collaboration}),\ }\href {\doibase 10.1140/epjc/s10052-006-0079-4} {\bibfield
   {journal} {\bibinfo  {journal} {Eur. Phys. J. C}\ }\textbf {\bibinfo
  {volume} {48}},\ \bibinfo {pages} {329} (\bibinfo {year} {2006})}\BibitemShut
  {NoStop}%
%%CITATION = NUCL-EX/0612012;%%
\bibitem [{\citenamefont {Abelev}\ \emph {et~al.}(2013)\citenamefont {Abelev}
  \emph {et~al.}}]{Abelev:2012ola}%
  \BibitemOpen
  \bibfield  {author} {\bibinfo {author} {\bibfnamefont {B.}~\bibnamefont
  {Abelev}} \emph {et~al.} (\bibinfo {collaboration} {ALICE Collaboration}),\
  }\href {\doibase 10.1016/j.physletb.2013.01.012} {\bibfield  {journal}
  {\bibinfo  {journal} {Phys. Lett. B}\ }\textbf {\bibinfo {volume} {719}},\
  \bibinfo {pages} {29} (\bibinfo {year} {2013})}\BibitemShut {NoStop}%
\bibitem [{\citenamefont {Aad}\ \emph {et~al.}(2013)\citenamefont {Aad} \emph
  {et~al.}}]{Aad:2013fja}%
  \BibitemOpen
  \bibfield  {author} {\bibinfo {author} {\bibfnamefont {G.}~\bibnamefont
  {Aad}} \emph {et~al.} (\bibinfo {collaboration} {ATLAS Collaboration}),\
  }\href {\doibase 10.1016/j.physletb.2013.06.057} {\bibfield  {journal}
  {\bibinfo  {journal} {Phys. Lett. B}\ }\textbf {\bibinfo {volume} {725}},\
  \bibinfo {pages} {60} (\bibinfo {year} {2013})}\BibitemShut {NoStop}%
\bibitem [{\citenamefont {Chatrchyan}\ \emph {et~al.}(2013)\citenamefont
  {Chatrchyan} \emph {et~al.}}]{Chatrchyan:2013nka}%
  \BibitemOpen
  \bibfield  {author} {\bibinfo {author} {\bibfnamefont {S.}~\bibnamefont
  {Chatrchyan}} \emph {et~al.} (\bibinfo {collaboration} {CMS Collaboration}),\
  }\href {\doibase 10.1016/j.physletb.2013.06.028} {\bibfield  {journal}
  {\bibinfo  {journal} {Phys. Lett. B}\ }\textbf {\bibinfo {volume} {724}},\
  \bibinfo {pages} {213} (\bibinfo {year} {2013})}\BibitemShut {NoStop}%
\bibitem [{\citenamefont {Adare}\ \emph
  {et~al.}(2013{\natexlab{b}})\citenamefont {Adare} \emph
  {et~al.}}]{Adare:2013piz}%
  \BibitemOpen
  \bibfield  {author} {\bibinfo {author} {\bibfnamefont {A.}~\bibnamefont
  {Adare}} \emph {et~al.} (\bibinfo {collaboration} {PHENIX Collaboration}),\
  }{\bibfield  {journal}
  {\bibinfo  {journal} {Phys. Rev. Lett.}\ }\textbf {\bibinfo {volume} {111}},\
  \bibinfo {pages} {212301} (\bibinfo {year} {2013}{\natexlab{b}})}\BibitemShut
  {NoStop}%
\bibitem [{\citenamefont {McGlinchey}\ \emph {et~al.}(2013)\citenamefont
  {McGlinchey}, \citenamefont {Frawley},\ and\ \citenamefont
  {Vogt}}]{McGlinchey:2012bp}%
  \BibitemOpen
  \bibfield  {author} {\bibinfo {author} {\bibfnamefont {D.}~\bibnamefont
  {McGlinchey}}, \bibinfo {author} {\bibfnamefont {A.~D.}\ \bibnamefont
  {Frawley}}, \ and\ \bibinfo {author} {\bibfnamefont {R.}~\bibnamefont
  {Vogt}},\ }\href {\doibase 10.1103/PhysRevC.87.054910} {\bibfield  {journal}
  {\bibinfo  {journal} {Phys. Rev. C}\ }\textbf {\bibinfo {volume} {87}},\
  \bibinfo {pages} {054910} (\bibinfo {year} {2013})}\BibitemShut {NoStop}%
\bibitem [{\citenamefont {Adare}\ \emph
  {et~al.}(2013{\natexlab{c}})\citenamefont {Adare} \emph
  {et~al.}}]{Adare:2013ezl}%
  \BibitemOpen
  \bibfield  {author} {\bibinfo {author} {\bibfnamefont {A.}~\bibnamefont
  {Adare}} \emph {et~al.} (\bibinfo {collaboration} {PHENIX Collaboration}),\
  }\href {\doibase 10.1103/PhysRevLett.111.202301} {\bibfield  {journal}
  {\bibinfo  {journal} {Phys. Rev. Lett.}\ }\textbf {\bibinfo {volume} {111}},\
  \bibinfo {pages} {202301} (\bibinfo {year} {2013}{\natexlab{c}})}\BibitemShut
  {NoStop}%
\bibitem [{\citenamefont {Winn}\ \emph {et~al.}()\citenamefont {Winn} \emph
  {et~al.}}]{Winn:HP2013}%
  \BibitemOpen
  \bibfield  {author} {\bibinfo {author} {\bibfnamefont {M.}~\bibnamefont
  {Winn}} \emph {et~al.},\ }\href@noop {} {}\bibinfo {note} {(ALICE
  Collaboration) unpublished}\BibitemShut {NoStop}%
\bibitem [{\citenamefont {Louren\c{c}o}\ \emph {et~al.}()\citenamefont
  {Louren\c{c}o}, \citenamefont {Vogt},\ and\ \citenamefont
  {W{\"{o}}hri}}]{Lourenco:2008sk}%
  \BibitemOpen
  \bibfield  {author} {\bibinfo {author} {\bibfnamefont {C.}~\bibnamefont
  {Louren\c{c}o}}, \bibinfo {author} {\bibfnamefont {R.}~\bibnamefont {Vogt}},
  \ and\ \bibinfo {author} {\bibfnamefont {H.~K.}\ \bibnamefont
  {W{\"{o}}hri}},\ }\href@noop {} {}\bibinfo {note} {J. High Energy Phys. {\bf
  02 (2009)} 014}\BibitemShut {NoStop}%
\bibitem [{\citenamefont {Arleo}\ and\ \citenamefont
  {Peigne}()}]{Arleo:2012rs}%
  \BibitemOpen
  \bibfield  {author} {\bibinfo {author} {\bibfnamefont {F.}~\bibnamefont
  {Arleo}}\ and\ \bibinfo {author} {\bibfnamefont {S.}~\bibnamefont {Peigne}},\
  }\href@noop {} {}\bibinfo {note} {ArXiv:1212.0434}\BibitemShut {NoStop}%
\bibitem [{\citenamefont {Arleo}\ \emph {et~al.}()\citenamefont {Arleo},
  \citenamefont {Kolevatov}, \citenamefont {Peigne},\ and\ \citenamefont
  {Rustamova}}]{Arleo:2013zua}%
  \BibitemOpen
  \bibfield  {author} {\bibinfo {author} {\bibfnamefont {F.}~\bibnamefont
  {Arleo}}, \bibinfo {author} {\bibfnamefont {R.}~\bibnamefont {Kolevatov}},
  \bibinfo {author} {\bibfnamefont {S.}~\bibnamefont {Peigne}}, \ and\ \bibinfo
  {author} {\bibfnamefont {M.}~\bibnamefont {Rustamova}},\ }\href@noop {}
  {}\bibinfo {note} {J. High Energy Phys. {\bf 05 (2013)} 155}\BibitemShut
  {NoStop}%
\bibitem [{\citenamefont {Kharzeev}\ and\ \citenamefont
  {Tuchin}(2006)}]{Kharzeev:2005}%
  \BibitemOpen
  \bibfield  {author} {\bibinfo {author} {\bibfnamefont {D.}~\bibnamefont
  {Kharzeev}}\ and\ \bibinfo {author} {\bibfnamefont {K.}~\bibnamefont
  {Tuchin}},\ }\href {\doibase 10.1016/j.nuclphysa.2006.01.017} {\bibfield
  {journal} {\bibinfo  {journal} {Nucl. Phys. A}\ }\textbf {\bibinfo {volume}
  {770}},\ \bibinfo {pages} {40} (\bibinfo {year} {2006})}\BibitemShut
  {NoStop}%
\bibitem [{\citenamefont {Chen}\ and\ \citenamefont {Ko}(2006)}]{Chen:2005zy}%
  \BibitemOpen
  \bibfield  {author} {\bibinfo {author} {\bibfnamefont {L.-W.}\ \bibnamefont
  {Chen}}\ and\ \bibinfo {author} {\bibfnamefont {C.~M.}\ \bibnamefont {Ko}},\
  }\href {\doibase 10.1103/PhysRevC.73.014906} {\bibfield  {journal} {\bibinfo
  {journal} {Phys. Rev. C}\ }\textbf {\bibinfo {volume} {73}},\ \bibinfo
  {pages} {014906} (\bibinfo {year} {2006})}\BibitemShut {NoStop}%
\bibitem [{\citenamefont {Digal}\ \emph {et~al.}(2012)\citenamefont {Digal},
  \citenamefont {Satz},\ and\ \citenamefont {Vogt}}]{Digal:2012}%
  \BibitemOpen
  \bibfield  {author} {\bibinfo {author} {\bibfnamefont {S.}~\bibnamefont
  {Digal}}, \bibinfo {author} {\bibfnamefont {H.}~\bibnamefont {Satz}}, \ and\
  \bibinfo {author} {\bibfnamefont {R.}~\bibnamefont {Vogt}},\ }\href@noop {}
  {\bibfield  {journal} {\bibinfo  {journal} {Phys. Rev. C}\ }\textbf {\bibinfo
  {volume} {85}},\ \bibinfo {pages} {034906} (\bibinfo {year}
  {2012})}\BibitemShut {NoStop}%
%%CITATION = ARXIV:;%%
\bibitem [{\citenamefont {Akikawa}\ \emph {et~al.}(2003)\citenamefont {Akikawa}
  \emph {et~al.}}]{Akikawa:2005}%
  \BibitemOpen
  \bibfield  {author} {\bibinfo {author} {\bibfnamefont {H.}~\bibnamefont
  {Akikawa}} \emph {et~al.},\ }\href@noop {} {\bibfield  {journal} {\bibinfo
  {journal} {Nucl. Instrum. Methods A}\ }\textbf {\bibinfo {volume} {449}},\
  \bibinfo {pages} {537} (\bibinfo {year} {2003})}\BibitemShut {NoStop}%
\bibitem [{\citenamefont {Miller}\ \emph {et~al.}(2007)\citenamefont {Miller},
  \citenamefont {Reygers}, \citenamefont {Sanders},\ and\ \citenamefont
  {Steinberg}}]{Miller:2007ri}%
  \BibitemOpen
  \bibfield  {author} {\bibinfo {author} {\bibfnamefont {M.~L.}\ \bibnamefont
  {Miller}}, \bibinfo {author} {\bibfnamefont {K.}~\bibnamefont {Reygers}},
  \bibinfo {author} {\bibfnamefont {S.~J.}\ \bibnamefont {Sanders}}, \ and\
  \bibinfo {author} {\bibfnamefont {P.}~\bibnamefont {Steinberg}},\ }\href
  {\doibase 10.1146/annurev.nucl.57.090506.123020} {\bibfield  {journal}
  {\bibinfo  {journal} {Ann. Rev. Nucl. Part. Sci.}\ }\textbf {\bibinfo
  {volume} {57}},\ \bibinfo {pages} {205} (\bibinfo {year} {2007})}\BibitemShut
  {NoStop}%
%%CITATION = NUCL-EX/0701025;%%
\bibitem [{\citenamefont {Adare}\ \emph
  {et~al.}(2008{\natexlab{b}})\citenamefont {Adare} \emph
  {et~al.}}]{Adare:2007gn}%
  \BibitemOpen
  \bibfield  {author} {\bibinfo {author} {\bibfnamefont {A.}~\bibnamefont
  {Adare}} \emph {et~al.} (\bibinfo {collaboration} {PHENIX Collaboration}),\
  }\href {\doibase 10.1103/PhysRevC.77.024912, 10.1103/PhysRevC.79.059901}
  {\bibfield  {journal} {\bibinfo  {journal} {Phys. Rev. C}\ }\textbf {\bibinfo
  {volume} {77}},\ \bibinfo {pages} {024912} (\bibinfo {year}
  {2008}{\natexlab{b}})}\BibitemShut {NoStop}%
\bibitem [{\citenamefont {Sjostrand}\ \emph {et~al.}()\citenamefont
  {Sjostrand}, \citenamefont {Mrenna},\ and\ \citenamefont
  {Skands}}]{Sjostrand:pythia}%
  \BibitemOpen
  \bibfield  {author} {\bibinfo {author} {\bibfnamefont {T.}~\bibnamefont
  {Sjostrand}}, \bibinfo {author} {\bibfnamefont {S.}~\bibnamefont {Mrenna}}, \
  and\ \bibinfo {author} {\bibfnamefont {P.}~\bibnamefont {Skands}},\
  }\href@noop {} {\enquote {\bibinfo {title} {Pythia 6.4 physics and manual},}\
  }\bibinfo {note} {J. High Energy Phys. {\bf 05 (2006)} 026}\BibitemShut
  {NoStop}%
\bibitem [{\citenamefont {{\it GEANT 3.2.1}}(1994)}]{Brun:geant}%
  \BibitemOpen
  \bibfield  {author} {\bibinfo {author} {\bibnamefont {{\it GEANT 3.2.1}}},\
  }\href@noop {} {\emph {\bibinfo {title} {GEANT 3.2.1}}},\ \bibinfo
  {organization} {CERN Program Library} (\bibinfo {year} {1994}),\ \bibinfo
  {note}
  {\url{http://wwwasdoc.web.cern.ch/wwwasdoc/pdfdir/geant.pdf}}\BibitemShut
  {NoStop}%
\bibitem [{\citenamefont {Adare}\ \emph
  {et~al.}(2012{\natexlab{b}})\citenamefont {Adare} \emph
  {et~al.}}]{Adare:2012wf}%
  \BibitemOpen
  \bibfield  {author} {\bibinfo {author} {\bibfnamefont {A.}~\bibnamefont
  {Adare}} \emph {et~al.} (\bibinfo {collaboration} {PHENIX Collaboration}),\
  }\href {\doibase 10.1103/PhysRevC.86.064901} {\bibfield  {journal} {\bibinfo
  {journal} {Phys. Rev. C}\ }\textbf {\bibinfo {volume} {86}},\ \bibinfo
  {pages} {064901} (\bibinfo {year} {2012}{\natexlab{b}})}\BibitemShut
  {NoStop}%
\bibitem [{\citenamefont {Nagle}\ \emph {et~al.}(2011)\citenamefont {Nagle},
  \citenamefont {Frawley}, \citenamefont {Linden~Levy},\ and\ \citenamefont
  {Wysocki}}]{Nagle:2010ix}%
  \BibitemOpen
  \bibfield  {author} {\bibinfo {author} {\bibfnamefont {J.~L.}\ \bibnamefont
  {Nagle}}, \bibinfo {author} {\bibfnamefont {A.~D.}\ \bibnamefont {Frawley}},
  \bibinfo {author} {\bibfnamefont {L.~A.}\ \bibnamefont {Linden~Levy}}, \ and\
  \bibinfo {author} {\bibfnamefont {M.~G.}\ \bibnamefont {Wysocki}},\ }\href
  {\doibase 10.1103/PhysRevC.84.044911} {\bibfield  {journal} {\bibinfo
  {journal} {Phys. Rev. C}\ }\textbf {\bibinfo {volume} {84}},\ \bibinfo
  {pages} {044911} (\bibinfo {year} {2011})}\BibitemShut {NoStop}%
%%CITATION = ARXIV:1011.4534;%%
\bibitem [{\citenamefont {Eskola}\ \emph {et~al.}()\citenamefont {Eskola},
  \citenamefont {Paukkunen},\ and\ \citenamefont {Salgado}}]{Eskola:2009uj}%
  \BibitemOpen
  \bibfield  {author} {\bibinfo {author} {\bibfnamefont {K.~J.}\ \bibnamefont
  {Eskola}}, \bibinfo {author} {\bibfnamefont {H.}~\bibnamefont {Paukkunen}}, \
  and\ \bibinfo {author} {\bibfnamefont {C.~A.}\ \bibnamefont {Salgado}},\
  }\href@noop {} {}\bibinfo {note} {J. High Energy Phys. {\bf 04 (2009)}
  065}\BibitemShut {NoStop}%
\bibitem [{\citenamefont {Zhao}\ and\ \citenamefont
  {Rapp}(2010)}]{Zhao:2010nk}%
  \BibitemOpen
  \bibfield  {author} {\bibinfo {author} {\bibfnamefont {X.}~\bibnamefont
  {Zhao}}\ and\ \bibinfo {author} {\bibfnamefont {R.}~\bibnamefont {Rapp}},\
  }\href {\doibase 10.1103/PhysRevC.82.064905} {\bibfield  {journal} {\bibinfo
  {journal} {Phys. Rev. C}\ }\textbf {\bibinfo {volume} {82}},\ \bibinfo
  {pages} {064905} (\bibinfo {year} {2010})}\BibitemShut {NoStop}%
\end{thebibliography}


\end{document}